\documentstyle[12pt,epsf,axodraw]{article}

\def\mypagenumber{1}
\def\mydate{April 15, 1997}
\def\myend{\end{document}}

\def\Journal#1#2#3#4{{#1}{\bf #2} (#4) #3}

\def\NPB{{\em Nucl.\ Phys.} B}
\def\PLB{{\em Phys.\ Lett.} B}
\def\PRL{\em Phys.\ Rev.\ Lett. }
\def\PRB{{\em Phys.\ Rev.} B}
\def\PRD{{\em Phys.\ Rev.} D}
\def\PTP{{\em Prog.\ Theoret.\ Phys.} }
\def\AP{{\em Ann.\ Phys.\ (N.Y.)} }
\def\RMP{{\em Rev.\ Mod.\ Phys.} }

\def\CMP{\em Comm.\ Math.\ Phys. }
\def\MPLA{{\em Mod.\ Phys.\ Lett.} A}
\def\IJMPB{{\em Int.\ J.\ Mod.\ Phys.} B}
\def\IJMPA{{\em Int.\ J.\ Mod.\ Phys.} A}

\normalsize
\let\oldtheequation=\theequation
\def\doteqs#1{\setcounter{equation}{0}
            \def\theequation{{#1}.\oldtheequation}}
\newcounter{sxn}
\def\sx#1{\addtocounter{sxn}{1} \bigskip\medskip \goodbreak
\noindent{\bf\leftline{\thesxn.~~#1}} \nobreak \medskip}
\def\sxn#1{\sx{#1} \doteqs{\thesxn}}

\newcounter{axn}
\def\ax#1{\addtocounter{axn}{1} \bigskip\medskip\goodbreak 
\noindent{\bf Appendix {\Alph{axn}.~~#1}} \nobreak \medskip}
\def\axn#1{\ax{#1} \doteqs{\Alph{axn}}}

\date{}

\newdimen\mybaselineskip
\newcommand{\beeq}{\begin{equation}}
\newcommand{\eneq}{\end{equation}}
\newcommand{\beqn}{\begin{eqnarray}}
\newcommand{\eeqn}{\end{eqnarray}}
\newcommand{\bpic}{\begin{picture}}
\newcommand{\epic}{\end{picture}}

\newcommand{\hh}{\;\;}

\def\mybig{\displaystyle \strut }

\def\d{\partial}
\def\dd{\partial}
\def\la{\raise.16ex\hbox{$\langle$} \, }
\def\ra{\, \raise.16ex\hbox{$\rangle$} }
\def\go{\rightarrow}

\def\next{{~~~,~~~}}

\def\onehalf{ \hbox{${1\over 2}$} }
\def\threehalves{ \hbox{${3\over 2}$} }

\def\psibar{ \psi \kern-.65em\raise.6em\hbox{$-$} }
\def\mbar{ m \kern-.78em\raise.4em\hbox{$-$}\lower.4em\hbox{} }

\def\Abar{{\overline A}}
\def\Bbar{{\overline B}}
\def\Cbar{{\overline C}}
\def\Dbar{{\overline D}}
\def\MSbar{{\overline {\rm MS}}}

\def\tM{{\tilde M}}
\def\tI{{\widetilde I}}

\def\L{ {\cal L} }
\def\O{{\rm O}}
\def\ep{\epsilon}
\def\eps{\epsilon^{\mu\nu\rho}}

\def\vphi{ {\varphi} }
\def\hg{{\hat g}}
\def\hp{{\hat p}}
\def\hq{{\hat q}}

\def\eff{{\rm eff}}
\def\min{{\rm min}}
\def\div{{\rm div}}
\def\rint{{\rm int}}
\def\CS{{\rm CS}}
\def\CW{{\rm CW}}
\def\Hi{{\rm H}}

\def\tma{{\tilde m_1}}
\def\tmb{{\tilde m_2}}
\def\tmc{{\tilde m_3}}

\def\myfrac#1#2{{\mybig #1\over \mybig #2}}

\def\n@space{\nulldelimiterspace=0pt \mathsurround=0pt }
\def\huge#1{{\hbox{$\left#1\vbox to 20.5pt{}\right.\n@space$}}}

\def\myskip{\noalign{\kern 10pt}}
\def\myeqspace{\noalign{\kern 10pt}}
\def\crn{\cr\myeqspace}

\def\boxit#1{$\vcenter{\hrule\hbox{\vrule\kern3pt
    \vbox{\kern3pt\hbox{#1}\kern3pt}\kern3pt\vrule}\hrule}$}
\def\bigbox#1{$\vcenter{\hrule\hbox{\vrule\kern5pt
     \vbox{\kern5pt\hbox{#1}\kern5pt}\kern5pt\vrule}\hrule}$}

\def\ignore#1{{}}

\begin{document}

\bibliographystyle{unsrt}
\footskip 1.0cm

\thispagestyle{empty}
\setcounter{page}{\mypagenumber}

{\baselineskip=10pt \parindent=0pt \small
 \mydate 
\hfill \hbox{\vtop{\hsize=3.2cm   UMN-TH-1529/97\\  hep-th/9703121\\}}
%\vspace{6mm}
%\rightline{\mydate}
}

\vspace*{25mm}

%\centerline {\Large\bf  Two Loop Analysis of}
%\vspace*{8mm}
\centerline{\Large\bf Maxwell-Chern-Simons Scalar Electrodynamics}
\vspace*{8mm}
\centerline {\Large\bf  at Two Loop }
\vspace*{20mm}
\centerline{\large Pang-Ning Tan\footnote{e-mail:~ ptan@mnhepo.hep.umn.edu},
 Bayram Tekin\footnote{e-mail:~ tekin@mnhepo.hep.umn.edu},
and Yutaka Hosotani\footnote{e-mail:~ yutaka@mnhepw.hep.umn.edu}}

\vspace*{5mm}

\baselineskip=15pt
\centerline {\it School of Physics and Astronomy, University
       of  Minnesota}
\centerline {\it Minneapolis, Minnesota 55455, U.S.A.}

\vspace*{15mm}
%\baselinestretch{2.0}

%\normalsize

\begin{abstract}
\baselineskip=18pt
The Maxwell-Chern-Simons gauge theory with charged scalar fields is 
analyzed at two loop level.  The effective potential for the scalar fields
is derived in the closed form, and studied both analytically and
numerically.  It is shown that the $U(1)$ symmetry is spontaneously
broken in the massless scalar theory. Dimensional 
transmutation takes place in the Coleman-Weinberg limit in which the 
Maxwell term vanishes.  We point out the subtlety in defining the pure 
Chern-Simons scalar electrodynamics
and show that the Coleman-Weinberg limit must be
taken after renormalization. Renormalization group analysis
of the effective potential is also given at two loop.
\end{abstract}

\vspace*{5mm}

\centerline{PACS: 11.30.Qc, 11.15.Ex, 11.10.Kk}

{\small
\centerline{Keywords:  Maxwell-Chern-Simons theory, symmetry breaking}
\centerline{Coleman-Weinberg mechanism} 
}

%\end{titlepage}
 
\newpage

%\setcounter{page}{1}

%\textheight=20cm
%\headsep=0.75cm
%\vsize=20cm

%%%%%%%%%%%%%%%%%%%%%%%%%%%%%%%%%%%%%%%%%%%%%%%%%%%%%%%%%%%%%%%%%
\normalsize
\baselineskip=22pt plus 1pt minus 1pt
\parindent=25pt
%\vspace*{5mm}

\sxn{Introduction}

In the previous paper we have evaluated  the effective potential of massless
scalar fields in three-dimensional $U(1)$ gauge theory to the two loop order
and have shown that  the $U(1)$ symmetry is spontaneously broken when the
Chern-Simons term is present for  gauge fields.\cite{Pang} In this paper we 
shall give
a full account of this theory, including the Coleman-Weinberg limit and
renormalization group analysis.  Subtlety in defining the 
Coleman-Weinberg limit is pointed out.  Numerical study of the two loop
effective potential is also presented.

There are many reasons for investigating three-dimensional $U(1)$ gauge theory
with both Maxwell and Chern-Simons terms.  Nonrelativistic Chern-Simons
theory serves as an effective theory of the quantum Hall system\cite{QHE}.
Chern-Simons interactions describe the change in statistics, and in general
fractional statistics.\cite{anyon} It was  argued that the system
of charged anyon gas leads to superconductivity, though experimental
support is lacking.\cite{super}

Relativistic three-dimensional gauge theory serves as an effective
theory  of
 four dimensional  theory at high temperature.
 In particular,  Maxwell-Chern-Simons
theory appears as an effective theory of QCD and the standard model of
electroweak  interactions. \cite{Linde,Kajantie,Efraty}

Maxwell-Chern-Simons theory has many unique features.  A photon acquires a 
topological
mass without breaking the gauge invariance.\cite{Deser,Schon}  When the 
$U(1)$ symmetry is spontaneously broken, photons appear with two different
masses.  In self-dual Chern-Simons theories many exact topological and
non-topological soliton solutions are available.\cite{Dunne}
In the Maxwell-Chern-Simons theory with Dirac fermions a  magnetic field can
be dynamically generated so that the Lorentz invariance is spontaneously
broken.\cite{Hosotani}  Pure non-Abelian Chern-Simons theory defines a
topological field theory,  playing an important role in the knot
theory.\cite{Witten}

Quantum aspect of the Maxwell-Chern-Simons gauge theory is under intense 
investigation in the literature. The Chern-Simons term is induced by 
Dirac fermions at one loop.\cite{DiracCS}    In non-Abelian theory the
Chern-Simons coefficient is quantized.\cite{Deser}  Non-Abelian gauge fields
themselves induce a  Chern-Simons term at one loop.\cite{Pisarski}
Pure non-Abelian Chern-Simons theory is expected to be
ultraviolet  finite.\cite{Martin}
The Coleman-Hill theorem assures that
corrections to the Chern-Simons coefficient are absent beyond one 
loop.\cite{Hill}
In the spontaneously broken non-Abelian gauge theory, however, corrections
could arise, depending on how the symmetry is broken.  In a certain type of
scalar field theory it has been argued that symmetry can be broken by
radiative corrections even at one loop.  In relativistic fermion theories
the resummation of ring diagrams  leads to spontaneous magnetization.
 \cite{Hosotani} Beta functions have been calculated in pure 
Chern-Simons gauge theories.\cite{Avdeev}

Yet, most arguments in the literature are limited to the one loop
approximation or the random phase approximation. One of the main concerns
in this paper is the phase structure, namely the symmetry structure, of
the scalar gauge theory particularly when the scalar fields are massless.
We shall show that one loop result is ambiguious, and one needs to go 
to two loop to find  definitive conclusions.

In this regard there is a subtle difference between the pure Chern-Simons
gauge theory and the Maxwell-Chern-Simons gauge theory.  Naively defined
in three dimensions, these theories have photon propagators which
behave, at large momenta, completely differently.  In the pure Chern-Simons
gauge theory the photon propagator behaves as $1/p$, whereas in the 
Maxwell-Chern-Simons gauge theory it behaves as $1/p^2$.  The ultraviolet
behavior is completely different.

This problem is tied to the renormalizability of
the theory.  First a regularization method must be specified which works
to all orders in perturbation theory. We adopt the dimensional regularization
method in the Maxwell-Chern-Simons theory.  The pure Chern-Simons theory
is defined in the limit of the vanishing Maxwell coefficient after
renormalization.  We show that the limit is well defined and exists only 
after renormalization.  

If the scalar fields self-interact only through  $\phi^6$ coupling in the 
pure Chern-Simons theory, the theory at the tree level does not
have any dimensional parameter. We define the pure Chern-Simons theory
in the manner described above, and show that the dimensional transmutation 
takes place at two loop.

Section 2 is devoted to the study of pure
complex scalar theory in 2+1 dimensions up to two loop. In section 3 we give
an analysis of super renormalizable real scalar $\lambda \,\phi^4$ theory.
Section 4 contains the definition of the gauge theory and the prescription
to dimensionally continue it to $n$ dimensions. One and two loop 
calculations are given in sections 5 and 6, respectively. In Section 6,  
the renormalized effective potential is given in the analytic
form in the limit of small and large scalar fields. In section
7 the Coleman-Weinberg limit of the effective potential is obtained. 
Section 8 includes an analysis of pure Maxwell theory, namely parity 
preserving 2+1 dimensional scalar QED. Renormalization 
for arbitrary value of the field is carried out 
numerically in section 9. Divergence structure of the theory by using
power counting method is discussed in section 10. We use the renormalization
group arguments to find the beta functions in section 11. Summary is 
given in section 12. Two loop
calculations are quite tedious. We have collected relevant integrals
in appendices.

\sxn{Pure Complex Scalar Theory}

In this section we analyze a complex scalar field theory in three
dimensions.  The most general renormalizable $U(1)$ invariant 
Lagrangian for $\Phi = (\phi_1 + i \phi_2)/\sqrt{2}$ is given by
\beeq
\L= {1\over 2} (\dd\phi_1)^2 + {1\over 2} (\dd\phi_2)^2 
 - {m^2\over 2} (\phi_1^2+\phi_2^2) 
 -{\lambda\over 4!} (\phi_1^2+\phi_2^2)^2
 -{\nu\over 6!} (\phi_1^2+\phi_2^2)^3 ~.
\label{Lagrangian1}
\eneq
The metric is given by $g^{\mu\nu}=diag(+,-,-)$.
When $m^2=\lambda=0$, the theory at the tree level does not contain
any dimensional parameter.  At the quantum level, however, a dimensional
scale enters in the 
definition of the renormalized coupling constant $\nu$ and a question arises
whether or not the $U(1)$ symmetry is spontaneously broken by radiative 
corrections.  We shall show that at  two loop the effective potential is 
minimized at a non-vanishing $\phi$,
but the minimum occurs outside the region of the validity of the 
perturbation theory.  

We are going to evaluate the effective potential for (\ref{Lagrangian1}) 
for arbitrary values of the parameters $m^2$, $\lambda$, and $\nu$ at 
two loop.  Let us recall the general formula for the effective action.
For a Lagrangian $\L(\phi)$ the effective action $\Gamma(\vphi)$
is \cite{Jackiw}
\beqn
&&\Gamma[\vphi] = \int d^nx \, \L[\vphi] + \tilde \Gamma[\vphi] \cr
\noalign{\kern 5pt}
&&\tilde \Gamma[\vphi] = - i\hbar \, \ln \int {\cal D}\phi \,
\exp {i\over \hbar} \int d^nx \bigg[ 
  \onehalf \phi \, i D_F^{-1}(\vphi)\, \phi + \L_\rint(\phi;\vphi)
  - {\delta \tilde\Gamma(\vphi)\over \delta \vphi} \, \phi \bigg]
\label{effectivrAction1}
\eeqn
where
\beqn
&&\L[\phi+\vphi] = \L[\vphi] +{\delta\L(\vphi)\over \delta \vphi_a} \phi_a
+\onehalf \phi_a i[D_F^{-1}(\vphi)]^{ab} \phi_b + \L_\rint(\phi;\vphi)\cr
&& i[D_F^{-1}(\vphi)]^{ab} = 
  {\delta^2 \L(\vphi)\over \delta\vphi_a \delta\vphi_b}  ~~~.
\label{effectiveAction2}
\eeqn
The above matrix equation gives propagators of the theory.  
$\tilde\Gamma[\vphi]$ is the sum of one-particle irreducible diagrams.
At one loop
\beeq
\Gamma[\vphi]^{(1)} = \cases{
{\mybig i\hbar \over\mybig 2}  \, \ln \det [iD_F^{-1}(\vphi)]  
     &for bosons\cr\cr
-i \hbar \, \ln \det [iD_F^{-1}(\vphi)]  &for Dirac fields\cr}
\label{1loopAction}
\eneq
For constant $\vphi(x)=\vphi$ the effective potential is
\beqn
V_\eff(\vphi) &=& V^{({\rm tree})}(\vphi) \, + \,
{\hbar \over 2}  \int {d^nk\over i(2\pi)^n} \, 
   \ln \det [iD_F^{-1}(k;\vphi)] \cr
\myskip
&& + i\hbar \bigg\langle \exp \bigg( {i\over \hbar} \int d^nx \,
  \L_\rint(\phi,\vphi) \bigg) \bigg\rangle_{\rm 1PI} 
\label{effectivePotential}
\eeqn
where the propagator is written in the momentum space.

The Lagrangian (\ref{Lagrangian1}) becomes, 
after the shifting $\phi_1 \go v + \phi_1$,
\beqn
\L ~~ &=& \L_{(0)} + \cdots + \L_{(6)} \cr
\myskip
\L_{(0)} &=& -{m^2\over 2} v^2 -{\lambda\over 4!}v^4-{\nu\over 6!} v^6\cr
\myskip
\L_{(2)} &=& {1\over 2} (\dd\phi_1)^2 + {1\over 2} (\dd\phi_2)^2 
 - \onehalf m_1^2 \phi_1^2 - \onehalf m_2^2 \phi_2^2 \cr
\myskip
\L_{(3)} &=& -{\lambda\over 3!} v\phi_1 (\phi_1^2+\phi_2^2)
  - {\nu\over 6!} 4 v^3 \phi_1 (5\phi_1^2 + 3\phi_2^2)\cr
\myskip
\L_{(4)} &=& -{\lambda\over 4!} (\phi_1^2+\phi_2^2)^2
-{\nu\over 6!} \, 3v^2 (5\phi_1^2+\phi_2^2)(\phi_1^2+\phi_2^2) \cr
\myskip
\L_{(5)} &=& -{\nu\over 5!} \, v\phi_1 (\phi_1^2+\phi_2^2)^2\cr
\myskip
\L_{(6)} &=& - {\nu\over 6!} (\phi_1^2+\phi_2^2)^3
\eeqn
The linear term $\L_{(1)}$ may absorbed by the redefinition of the 
source and is irrelevant. The mass parameters are given by :
\beqn
m_1^2 &=& m_1(v)^2 = m^2 + {\lambda\over 2} \, 
	  v^2 + {\nu\over 24} \, v^4 \cr
\myskip
m_2^2 &=& m_2(v)^2 = m^2 + {\lambda\over 6} \, v^2 
	  + {\nu\over 120} \, v^4
\eeqn
In $n$ dimensional space-time the dimensions of the coupling constants and
fields are
\beeq
[m] = M \next [\lambda] = M^{4-n} 
\next [\nu] = M^{2(3-n)} \next [v^2] = M^{n-2} ~~~.
\eneq

The tree level effective potential is 
\beeq
V_\eff^{(0)} = 
{m^2\over 2} v^2 +{\lambda\over 4!}v^4 +{\nu\over 6!} v^6 + \Lambda
\label{Vtree}
\eneq
The last term in ({\ref{Vtree}}) is the cosmological constant which is
a function of the dimensional parameters. Although it is irrelevant for
the discussion of symmetry breaking, it plays an important role in 
renormalization group analysis.\cite{Bando}

The one loop effective potential is finite in the dimensional regularization
scheme and is given by  
\beqn
V_\eff^{(1)} &=& {\hbar\over 2} \int {d^np\over i(2\pi)^n} \,
\bigg\{ ~ \ln\, \big[p^2 - m_1^2\big] + \ln\, \big[p^2 - m_2^2\big] 
\bigg\} \cr
\myskip
&=& - {\hbar\over 2} {\Gamma (- {n\over 2}) \over
      (4 \pi)^{n\over 2}} \bigg [ m_1^n + m_2^n \bigg] \cr
\myskip
&=& -{\hbar\over 12\pi} \, \mu^{n-3} ( 
m_1^3 + m_2^3 )+ {\rm O}(n-3)~~~.
\label{1loopV1}
\eeqn

At two loop, we denote
\beqn
&&\L_{(3)} = - \beta_1 \phi_1^3 - \beta_2 \phi_1 \phi_2^2  \cr \myskip
&& \beta_1 = {\lambda\over 3!} v + {\nu\over 36} v^3 \next 
   \beta_2 = {\lambda\over 3!} v + {\nu\over 60} v^3 \cr
\myskip
&&\L_{(4)} = - \alpha_1 \phi_1^4 - \alpha_2 \phi_2^4 
             - \alpha_3 \phi_1^2\phi_2^2 \cr 
\myskip
&&\alpha_1 = {\lambda\over 4!} + {\nu\over 2\cdot 4!} v^2 \next
  \alpha_2 = {\lambda\over 4!} + {\nu\over 2\cdot 5!} v^2 \next
  \alpha_3 = {2\lambda\over 4!} + {3\nu\over 5!} v^2 
\eeqn
The cubic part of the Lagrangian gives rise to theta shape diagrams which 
are reduced to the following.

\begin{center}
\bpic(25, 0)
\put(18,3){\circle{36}}
\put(0,3){\line(1,0){36}}
\epic
\end{center}
\vskip 0.2 cm
\beqn
I( m_a,m_b,m_c;n) &\equiv& \int \frac{d^nqd^nk}{(2\pi)^{2n}}
   \frac{1}{\left[ (q+k)^2+m_a^2\right] (q^2+m_b^2)(k^2+m_c^2)}\crn
&=& I(m_b, m_a,m_c;n) \quad {\rm etc.} \crn
&=& I^{\rm div} + \tI (m_a+m_b+m_c)  \crn
I^{\rm div} &=&
{\mu^{2(n-3)}\over 32 \pi^2}  \bigg\{ - {1\over n-3} - \gamma_E + 1 
   + \ln 4\pi  \bigg\} \crn
 \tI (m) &=& -
{\mu^{2(n-3)}\over 16 \pi^2}  \ln {m\over  \mu} ~~~.
\label{2loopIntegral0}
\eeqn
The derivation of (\ref{2loopIntegral0}) is given in  Appendix B.  We have
split $I$ function into divergent and finite parts for later convenience. 
The quartic part of the Lagrangian produces  two loop diagrams which are
reduced to the integral
\beqn
J( m_a,m_b;n) &\equiv& \int \frac{d^nqd^nk}{(2\pi)^{2n}}
    \frac{1}{(q^2+m_a^2)(k^2+m_b^2)} \crn
&=&
{\mu^{2(n-3)}\over 16 \pi^2} m_a \, m_b \bigg [ 1 + (n-3) 
	\psi \big( -{1\over 2} \big) + (n-3) \ln \, \bigg (
	{m_a \, m_b \over 4\pi \mu^2} \bigg ) \crn
&& \hskip 3 cm + {\rm O}(n - 3)^2 \bigg ] 
\eeqn
Therefore, two loop contributions  are
\beqn
V_\eff^{(2)} &=& {-\hbar^2 \over 2} \bigg\{ 6 \beta_1^2 \,  I(m_1,m_1,m_1) 
        +  2 \beta_2^2 \,  I(m_1,m_2,m_2) \bigg\} \cr
\myskip
&& +  \hbar^2  \bigg\{ 3 \alpha_1 J(m_1, m_1) + 3 \alpha_2 J(m_2, m_2)
   + \alpha_3 J(m_1, m_2) \bigg\} \crn
&=&  - \hbar^2 \, \bigg [ 3 \beta_1^2 + \beta_2^2 \bigg ]\, I^\div \cr
\myskip
&&  + {\mu^{2(n-3)}\hbar^2 \over 32\pi^2} 3 \beta_1^2 
	\ln { 9 \, m_1^2 \over \mu^2} 
    + {\mu^{2(n-3)}\hbar^2 \over 32\pi^2} \beta_2^2 
	\ln {(m_1 + 2 m_2)^2 \over \mu^2} \cr
\myskip
&& + {\mu^{2(n-3)}\hbar^2 \over 16\pi^2} \bigg \{
       3 \alpha_1 \, m_1^2 + 3 \alpha_2 \, m_2^2 + \alpha_3 m_1 \, m_2
       \bigg \} ~~.
\label{2loopV1}
\eeqn

Combining (\ref{Vtree}), (\ref{1loopV1}), (\ref{2loopV1}), one finds 
the total effective potential to O($\hbar^2$) is, up to counter terms,
\beqn
V_\eff(v;n) &=& {1\over 2} m^2 v^2 + {\lambda \over 4!} v^4 
	+ {\nu \over 6!} v^6 + \Lambda
	- {\hbar \over 12 \pi} \mu^{n-3}
	   \bigg( m_1^3 + m_2^3 \bigg) \cr
\myskip
&& - \hbar^2 
    \bigg [ \bigg( {\lambda \over 6} v + {\nu \over 60} v^3 \bigg)^2
       + 3 \bigg( {\lambda \over 6} v + {\nu \over 36} v^3 \bigg)^2
    \bigg ] \, I^\div  \cr
\myskip
&& + {\hbar^2 \over 32\pi^2} \mu^{2(n-3)} \bigg \{
     \bigg( {\lambda \over 6} v + {\nu \over 60} v^3 \bigg)^2 
	\ln {(m_1 + 2 m_2)^2 \over \mu^2} 
   + 3 \bigg( {\lambda \over 6} v + {\nu \over 36} v^3 \bigg)^2
	\ln {9 m_1^2 \over \mu^2} \bigg\} \cr
\myskip
&& + {\hbar^2 \over 16\pi^2} \mu^{2(n-3)} \bigg \{
     3 \bigg( {\lambda \over 4!} + {15\nu \over 6!} v^2 \bigg) m_1^2
   + 3 \bigg( {\lambda \over 4!} + {3 \nu \over 6!} v^2 \bigg) m_2^2 \cr
\myskip
&& \hskip 2 cm
   + 2 \bigg( {\lambda \over 4!} + {9 \nu \over 6!} v^2 \bigg) m_1 \, m_2
    \bigg\} ~~~.
\label{barePotential}
\eeqn

Beta functions for various coupling constants can be found in  variety 
of ways.  One way is  to evaluate corresponding Feynman 
diagrams to find divergent parts or counter terms.  An alternative way,
which is suited in our approach, is to find beta functions from
the requirement that 
the effective potential satisfy the renormalization group equation.
Both methods must give the same result.
We shall show in the rest of this section that both methods yield the same beta
functions in the pure scalar theory.

First we write down the renormalization group equation satisfied by
the effective potential in the $\MSbar$ scheme.   The $\MSbar$ regularization
scheme consists of absorbing terms proportional to
$-(n-3)^{-1} - \gamma_{\rm E} + 1 + \ln 4\pi$ by counter terms.  The resultant
effective potential  $V_\eff(v)^\MSbar$ obtained from (\ref{barePotential})  is
finite.  As the bare theory is independent of the dimensional parameter $\mu$,
it obeys
\beeq
\bigg [ \mu {\dd \over \dd \mu} + \beta_{m^2} {\dd \over \dd {m^2}} 
+ \beta_{\lambda} {\dd \over \dd \lambda} 
+ \beta_{\nu} {\dd \over \dd \nu} + \beta_{\Lambda} {\dd \over \dd \Lambda} 
- \gamma_{\phi} \, v {\dd \over \dd v} \bigg ] 
V_\eff(v)^\MSbar = 0
\label{RGeq1}
\eneq
where the beta functions and anomalous dimension are defined by
\beqn
&&\beta_{\lambda} = \mu {\dd\lambda \over \dd\mu},   \hskip 2cm
  \beta_{m^2} = \mu {\dd{m^2} \over \dd\mu} \cr
\myskip
&&\beta_{\nu} = \mu {\dd\nu \over \dd\mu},  \hskip 2 cm
\beta_{\Lambda} = \mu {\dd\Lambda \over \dd\mu} \cr
\myskip
&& \gamma_{\phi} = {\mu \over 2} {\dd\ln Z_\phi \over \dd\mu}  ~~.
\label{beta-fun1}
\eeqn

In the $\MSbar$ scheme, the $\beta$'s and $\gamma_\phi$ are functions of
various coupling constants and $\hbar$.  Eq.\ (\ref{RGeq1}) is an exact
relation, and is valid for arbitrary $v$ and to each  order in $\hbar$.
As can be easily shown, the anomalous dimension $\gamma_\phi$ vanishes
up to two loop, or to O($\hbar^2$).  
To O$(\hbar)$
\beeq
\beta_{m^2}^{(1)} {v^2 \over 2}  + \beta_{\lambda}^{(1)} {v^4 \over 4!}
+ \beta_{\nu}^{(1)} {v^6 \over 6!} + \beta_{\Lambda}^{(1)}  = 0~.
\label{RGeq2}
\eneq
Hence
\beeq
\beta_{m^2}^{(1)} = \beta_{\lambda}^{(1)} = \beta_{\nu}^{(1)} =
\beta_{\Lambda}^{(1)} = 0 ~~.
\label{beta-fun2}
\eneq

To O$(\hbar^2)$, Eq.\ (\ref{RGeq1}) becomes
\beqn
&& - {\hbar^2 \over 16 \pi^2} \Bigg\{ 
      \bigg( {\lambda \over 6} v + {\nu \over 60} v^3 \bigg)^2 
  + 3 \bigg( {\lambda \over 6} v + {\nu \over 36} v^3 \bigg)^2 \Bigg\} \cr
\myskip
&&\hskip 1cm  
+ \beta_{m^2}^{(2)} {v^2 \over 2} + + \beta_{\lambda}^{(2)} {v^4 \over 4!}
+ \beta_{\nu}^{(2)} {v^6 \over 6!} + \beta_{\Lambda}^{(2)} = 0 ~~.
\label{RGeq3}
\eeqn
Comparing the coefficients term by term, we find 
\beqn
&& \beta_{\Lambda}^{(2)} = 0 \next
\hskip 2cm
   \beta_{m^2}^{(2)} = {\hbar^2 \over 72 \pi^2} \lambda^2 \cr
\myskip
&& \beta_{\lambda}^{(2)} = {\hbar^2 \over 20 \pi^2} \lambda \nu \next
\hskip 1cm
   \beta_{\nu}^{(2)} =  {7 \hbar^2 \over 60 \pi^2} \nu^2 
\label{beta_purescalar}
\eeqn

The same result is obtained by the 
 conventional method of finding beta functions. The superficial degree 
of divergence $\omega$ for a given Feynman diagram is 
\beeq
\omega = 3 - {E\over 2} - V_4
\label{superficialdeg}
\eneq
where $V_4$ refers to the number of vertices of quartic coupling while $E$ 
is the number of the external lines. 

For diagrams contributing to $\beta_{m^2}$,  $E$=2 and $V_4$=2  to
O($\hbar^2$). There are two divergent diagrams   of the  form

\begin{center}
\bpic(70,12)
\put(20,3){\circle{36}}
\Line(-16,3)(56,3)
\epic
\end{center}
\vskip 0.5 cm
The self-energy term for $\phi_1$ (in $D^{-1} = p^2 - m_0^2 - \Sigma$) is
\beeq
 \Sigma(p) =  -\hbar^2 \lambda^2 \bigg\{ {1 \over 6} I(m_1, m_1, m_1)
              + {1 \over 18} I(m_1, m_2, m_2) \bigg\}  + {\rm O}(p^2)~~~.
\label{self-energy1}
\eneq
The O$(p^2)$ term is finite.   To this order the bare mass is
\beqn
m_0^2 &=& m^2 - \Sigma(0)^{\rm div} 
= m^2 + {2\hbar^2 \lambda^2\over 9} I^{\rm div}  ~~~,
\label{baremass}
\eeqn
where $I^{\rm div}$ is defined in (\ref{2loopIntegral0}).
The bare mass does not depend on $\mu$, and $\mu(d/d\mu) \lambda =
{\rm O}(\hbar^2)$.  Hence to O($\hbar^2$)
\beeq
\beta_{m^2} = \mu {d\over d\mu} m(\mu)^2 
= - {2\hbar^2 \lambda^2\over 9} \mu {d\over d\mu} I^{\rm div}
\Bigg|_{n=3}
= {\hbar^2\lambda^2\over 72\pi^2} ~~, 
\label{beta-fun3}
\eneq
which agrees with (\ref{beta_purescalar}).

Two loop diagrams contributing to $\beta_{\lambda}$ must have  $E$=4,
$V_4$=1, and $V_6$=1,  taking the form of 

%\vskip 0.5
%\hskip 3cm
\begin{center}
\bpic(70,12)
\put(20,3){\circle{36}}
\Line(-16,3)(56,3)
\Line(38,3)(56,11)
\Line(38,3)(56,-5)
\epic
\end{center}
\vskip 0.5 cm
The total vertex at zero momentum is
\beeq 
\lambda_0   -  \hbar^2  \bigg\{
	       {2\lambda \nu \over 3}\,  I(m_1, m_1, m_1)
             + {2\lambda \nu \over 15}\, I(m_1, m_2, m_2) \bigg\}  ~.
\eneq
The bare coupling constant is
\beeq
\lambda_0 = \lambda + {4\over 5} \hbar^2 \lambda \nu I^{\rm div} ~~.
\label{bare-lambda}
\eneq
The same result for $\beta_\lambda$ as in (\ref{beta_purescalar}) follows 
from $\mu (d/d\mu) \lambda_0 = 0$. 

Similarly, for ${\beta_{\nu}}$ 
there are two diagrams to be considered:
\begin{center}
\bpic(70,12)
\put(20,3){\circle{36}}
\Line(-16,3)(56,3)
\Line(-16,-5)(2,3)
\Line(-16,11)(2,3)
\Line(38,3)(56,11)
\Line(38,3)(56,-5)
\epic
\end{center} 
\vskip 0.5 cm
When all external lines are $\phi_1$ fields, the six point vertex at
zero momenta is 
\beeq 
{\nu_0 } - \hbar^2 \bigg\{ {5\nu^2 \over 3} I(m_1, m_1, m_1)
         +{\nu^2 \over 5}  I(m_1, m_2, m_2) \bigg\} 
\eneq
so that 
\beeq
\nu_0 = \nu + {28\over 15} \hbar^2 \nu^2 I^{\rm div} ~~~.
\label{bare-nu}
\eneq
$\mu (d/d\mu) \nu_0=0$ leads to the previous result for $\beta_\nu$ 
in (\ref{beta_purescalar}).

\def\hh{\hskip .5cm}

Now, consider the special case $m^2=\lambda=0$ i.e. when no dimensionful 
parameter appears at the tree level. 
In the $\MSbar$ renormalization scheme, the total
effective potential to O($\hbar^2$)  takes the form
\beeq
V_\eff(v;n) = {\Abar v^2\over 2} + {\Bbar v^4\over 4!} 
 + {\Cbar v^6\over 6!} 
   + {\Dbar v^6\over 6!} \ln {\mu^{2(3-n)} \nu v^4\over 4\pi\mu^2}  ~~.
\label{scalarPotential0}
\eneq
We impose the following renormalization conditions at $n=3$:
\beqn
 \, \, V_\eff \Bigg|_{v=0} \hskip .7cm  & =& 0 \cr \myskip
{\dd^2 V_\eff\over \dd v^2} \Bigg|_{v=0}\hh &=& m^2 = 0 \cr \myskip
{\dd^4 V_\eff\over \dd v^4} \Bigg|_{v=0}\hh &=& \lambda = 0 \cr \myskip
{\dd^6 V_\eff\over \dd v^6} \Bigg|_{v=M^{1/2}}&=& \nu(M) = \nu 
\label{rencond}
\eeqn
Note that $\nu(M)$ has to be defined at $M\not= 0$, as the effective
potential has a $\ln v$ singularity at $v=0$.  The resultant
effective potential is 
\beeq
V_\eff(v)^{\rm pure~scalar} = {1\over 6!} \, \nu(M) \, v^6
+ {1\over 6!} \, {7\hbar^2\over 120\pi^2} \, \nu(M)^2 \, v^6 
 \bigg( \ln {v^4\over M^2} - {49\over 5} \bigg) ~.
\label{scalarPotential1}
\eneq

At first glance, it seems that the potential has a minimum at
a nonvanishing $v \equiv v_{\rm min}$.  However,
 \beeq
\nu \ln v_{\rm min}^4/M^2 =  - {120\pi^2 \over 7\hbar^2} + {137 \over 15} 
\, \nu . \label{minCondition}
\eneq
For small $\nu$, the first term dominates and has an absolute value
much bigger  than one.  Since higher order corrections  produce higher
powers of ({\ref{minCondition}}), we conclude that the location of the new 
minimum occurs outside the domain of validity of perturbation theory, 
as in the Coleman-Weinberg limit of the 3+1 dimensional pure scalar theory.
One cannot draw any definitive conclusion concerning the symmetry
breaking from the above perturbative analysis.

\sxn{Real Scalar Theory}

In this section we analyze the 2+1 dimensional $\lambda \phi^4$
theory with a vanishing $\phi^6$ coupling. $\lambda$ has dimension of mass 
so that the theory is 
super-renormalizable. By looking at the superficial degree of divergence 
one can find that $\beta_\lambda$ is zero  to all orders in 
perturbation theory. This is seen at two loop by letting $\nu$
 be equal to zero in the equation ({\ref{beta_purescalar}}). 
Beta functions at three loop are found in \cite{Mckeon}. 
The $Z_2$ symmetric version of this theory, namely real scalar theory, has 
also been studied  both at one
loop and in the Gaussian approximation which 
gives an upper bound for the effective potential. Here we would like to 
extend
the analysis to two loop. 

In the $Z_2$ symmetric case the $\MSbar$ renormalized potential is given 
by           
\beqn 
V_\eff ^\MSbar &=& {1\over 2} m^2 v^2 +{\lambda\over 24} v^4 - {\hbar 
\over{12 \pi}}{ ( m^2 +{\lambda \over 2} v^2 )}^{3 \over 2} \nonumber \\ 
        &&+ { {\hbar}^2\over {128 \pi^2}} \lambda (m^2+{\lambda\over 2}v^2) 
         +{\hbar^2 \over {384 \pi^2} } \lambda^2 v^2 
           \ln {9 (m^2+ {\lambda\over 2} v^2)\over \mu^2} 
\eeqn 
The parameters $m^2$ and $\lambda$ are finite but 
otherwise arbitrary. We renormalize by 
\beqn
V_\eff(0) = 0 ~~ , \hskip 0.5 cm  V_\eff^{(2)}(0) = m^2 ~~ ,\hskip 0.5 cm  
V_\eff^{(4)}(0) = \lambda ~~.
\label{normalize1} 
\eeqn
Then one obtains 
\beqn
V_\eff & = &{1\over 2}( m^2 +{\hbar\lambda m \over {8\pi}}) v^2  
          +{\lambda\over 24} v^4  - {\hbar \over{12\pi}}
{ ( m^2 +{\lambda \over 2} v^2 )}^{3 \over 2}\nonumber \\
        && + {\hbar^2 \lambda^2 v^2 \over {384 \pi^2} }
 \ln (1+ {\lambda\over 2 m^2 } v^2) + {\hbar\over {12\pi}} m^3   
+ {\hbar\lambda^2 v^4\over {128\pi m}} - {\hbar^2\lambda^3 v^4\over 
{768\pi^2 m^2}}
\eeqn

In fig.1 we have plotted  one loop 
and two loop results. In the figure, one can see that for small values of
$\lambda/m$, one loop and two loop potential are close to each other. As
$\lambda/m$ increases, they start to deviate from each other.

For small $\lambda/m$ the symmetry is unbroken.   
For $27.811 < \lambda/m < 29.541 $ the two-loop effective potential 
is minimized at a  non-vanishing $v$.
It becomes unbounded from below for  $\lambda/m > 29.541$.
However, the perturbation
theory breaks down for such a large coupling. 

It has been shown by Stevenson and by Olsen et al.\ that in the 
Gaussian approximation the symmetry is spontaneously broken if the 
coupling $\lambda/m$ becomes sufficiently large.\cite{Olsen}
Our result in perturbation theory is valid for small $\lambda/m$, 
and is consistent with the result in the Gaussian approximation.
[Note that $\lambda$ in \cite{Olsen} is not normalized by the condition 
(\ref{normalize1}).]

\begin{figure}[htb]
\epsfxsize=11cm  
\centerline{
\epsffile[47 195 474 501]{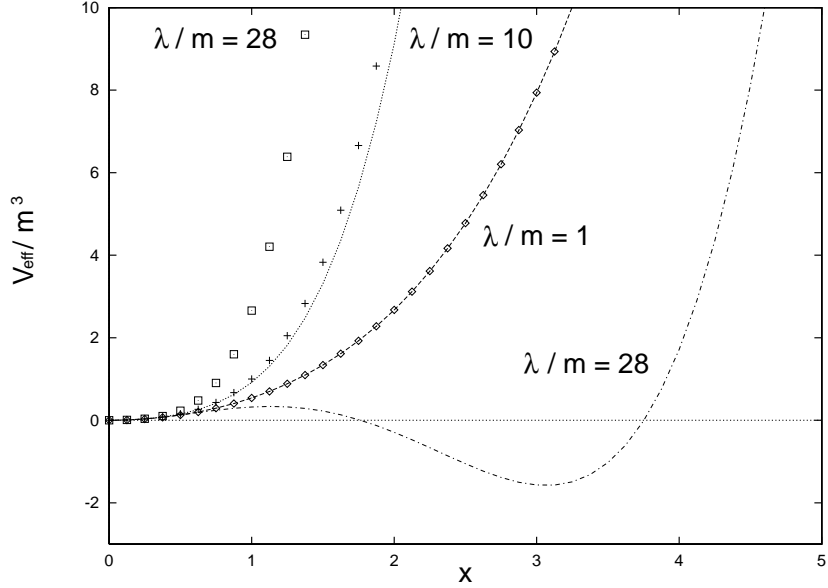}
}
%\vskip 1cm
\caption{Effective potential for $\phi^4$ real scalar theory using various
values of $\lambda / m$. One loop data are represented as points while two 
loop data are depicted as lines.}
\label{fig:1}
%\vskip 0.2cm
\end{figure}

\sxn{Gauge theory }

In the presence of $U(1)$ gauge interactions the  most general
renormalizable Lagrangian is given by 
\beqn
\L &=& - {a\over 4} F_{\mu\nu} F^{\mu\nu}
- {\kappa\over 2} \ep^{\mu\nu\rho} A_\mu \dd_\nu A_\rho
+ \L_{\rm g.f.} + \L_{\rm F.P.} \cr
\noalign{\kern 5pt}
&& +{1\over 2} (\dd_\mu\phi_1 - eA_\mu \phi_2)^2
 +{1\over 2} (\dd_\mu\phi_2 + eA_\mu \phi_1)^2 \cr
\noalign{\kern 5pt}
&&-{m^2\over 2} (\phi_1^2 +\phi_2^2) 
- {\lambda\over 4!}  (\phi_1^2 +\phi_2^2)^2 
-{\nu\over 6!}  (\phi_1^2 +\phi_2^2)^3 ~~. 
  \label{Lagrangian2}
\eeqn
In the $R_\xi$ gauge
\beqn
\L_{\rm g.f.}~ &=& - {1\over 2\alpha} (\dd_\mu A^\mu - \alpha ev \phi_2)^2 \cr
\noalign{\kern 5pt}
\L_{\rm F.P.} &=& - c^\dagger \, ( \dd^2 + \alpha e^2 v \phi_1) \, c 
\label{R-gauge}
\eeqn
We would like to find the effective potential $V_{\eff}[v]$ for the $\phi$
fields (say $\la \phi_1 \ra = v$, $\la \phi_2 \ra =0$) to the two loop
order.    In $n$ dimensions
\beqn
&&[\phi]=[A_\mu] = M^{(n-2)/2} \next
[a]=[\alpha]=M^0\cr
\noalign{\kern 5pt}
&&[e] = M^{(4-n)/2}  \next [\kappa] = M \cr
\noalign{\kern 5pt}
&&[m] = M \next  [\lambda] = M^{4-n}   \next  [\nu] = M^{2(3-n)} ~~.
    \label{dimension} 
\eeqn

Not all parameters in the Lagrangian (\ref{Lagrangian2}) are
independent.
By scaling $A_\mu' = t A_\mu$, one finds the equivalence relation
\beeq
(a, \kappa, e, \alpha) \sim
( {a\over t^2}, {\kappa\over t^2}, {e\over t}, t^2 \alpha)~,
\label{equivalence1}
\eneq
or
\beeq
(a, k={\kappa\over e^2},  e, \alpha) \sim
( {a\over t^2}, k,  {e\over t}, t^2 \alpha)~.
\label{equivalence2}
\eneq
Physics is independent of $t$.  If the renormalized $a=0$, physics 
in the  Landau gauge ($\alpha=0$) depends on $k=\kappa/e^2$, $m$,
$\lambda$,  and $\nu$.  In particular, with $m=\lambda=0$ the
classical theory contains no dimensional parameter.  As is shown
shortly, however, the $a=0$ theory should be defined by the limit
$a\go 0$.

After shifting $\phi_1 \go v + \phi_1$, the quadratic part of the 
Lagrangian (\ref{Lagrangian2}) is
\beqn
\L_{(2)} &=& 
{1\over 2} A_\mu ~ K^{\mu\nu} ~  A_\nu 
- c^\dagger \, ( \dd^2 + \alpha e^2 v^2) \, c
 - {1\over 2} \phi_1 (\dd^2 + m_1^2 ) \phi_1 
 - {1\over 2} \phi_2 (\dd^2 + m_2^2 ) \phi_2  \cr
\noalign{\kern 8pt}
K^{\mu\nu}  &=& 
\bigg\{  a  \dd^2  + (e v)^2 \bigg\}g^{\mu\nu} 
  - \Big( a - {1\over \alpha} \Big) \dd^\mu\dd^\nu  
+ \kappa \ep^{\mu\nu\rho} \dd_\rho  \cr
\noalign{\kern 8pt}
m_1^2 &=& m_1^2(v) = m^2 + {\lambda\over 2} v^2   + {\nu\over 24} v^4\cr
\noalign{\kern 8pt}
m_2^2 &=& m_2^2(v) =  m^2 + {\lambda\over 6} v^2 +  {\nu\over 120} v^4
   + \alpha (ev)^2
\label{Lagran0b}
\eeqn

In this paper, we adopt the dimensional regularization method. 
The definition of the totally antisymmetric tensor, $\eps$,  depends
on the three dimensionality of spacetime.   Below we define the $\eps$
tensor in $n$ dimensions in a way that it
stays  essentially in three 
dimensions.  This definition was initially proposed by  t'Hooft and
Veltman.\cite{t'Hooft}  It has
been shown that Slavnov-Taylor identities are satisfied with this 
definition,\cite{Chen} and that the Maxwell term improves the 
ultraviolet
behavior of the gauge field propagator.\cite{Martin}
  
In $n$ dimensions we define $\eps$ and $\hg^{\mu\nu}$ by 
\beqn 
\eps &=& \cases{
\pm 1 &if $(\mu,\nu,\rho)$= permutation of (0,1,2)\cr
0&otherwise.\cr}  \crn
\hg^{\mu\nu} &=& \cases{ +1&for $\mu=\nu=0$\cr
                       -1&for $\mu=\nu=1,2$\cr
                        0&otherwise.\cr}
\label{eps-3dmetric}
\eeqn
Then
\beqn
&&\ep^{\mu\nu\rho} {\ep^{\lambda\eta}}_\rho
 = \hg^{\mu\lambda} \hg^{\nu\eta} - \hg^{\mu\eta} \hg^{\nu\lambda}\cr
&&g^{\mu\nu} {\hg_\nu}^{~\lambda} = \hg^{\mu\lambda} 
\label{property1}
\eeqn
We denote $\hp^\mu = \hg^{\mu\nu} p_\nu$ etc.

The inverse of $K^{\mu\nu}$ in (\ref{Lagran0b}) is found easily.
In general
\beqn
&&K^{\mu\nu} = A g^{\mu\nu} + B p^\mu p^\nu - i\kappa \eps p_\rho\cr
\noalign{\kern 8pt}
&&K^{-1}_{\nu\lambda} =
{1\over A} \bigg( g_{\nu\lambda} 
      - {B\over A+p^2 B} \, p_\nu p_\lambda \bigg)
+ {\kappa^2\over A(A^2 - \kappa^2 \hp^2)} \,
       (\hp^2 \hg_{\nu\lambda} - \hp_\nu \hp_\lambda) \cr
\noalign{\kern 5pt}
&&\hskip 6cm
+ {i\kappa\over A^2 - \kappa^2 \hp^2} \, \ep_{\nu\lambda\rho} p^\rho.
\label{inverse1}
\eeqn
In our case
$A= -a p^2 + (ev)^2$, $B= a - \alpha^{-1}$ so that 
\beqn
K^{-1}_{\nu\lambda} &=&
-{1\over d(p^2)} \bigg( g_{\nu\lambda} 
 - (1-a\alpha)\, {p_\nu p_\lambda \over p^2 - \alpha (ev)^2}  \bigg) \cr
\noalign{\kern 8pt}
&&\hskip .5cm - {\kappa^2 \hp^2\over d(p^2) [d(p^2)^2  - \kappa^2 \hp^2]} \,
  \bigg(\hg_{\nu\lambda} - {\hp_\nu \hp_\lambda\over \hp^2} \bigg) 
    + {i\kappa \ep_{\nu\lambda\rho} p^\rho
   \over d(p^2)^2 - \kappa^2 \hp^2} \,  \cr
\noalign{\kern 14pt}
d(p^2)&=& a p^2 - (ev)^2 .
\label{inverse2}
\eeqn
In the Landau gauge,  $\alpha=0$, 
\beqn
K^{-1}_{\nu\lambda}\Big|_{\alpha=0}
&=& -{1\over d(p^2)} \bigg( g_{\nu\lambda} 
 -  {p_\nu p_\lambda \over p^2 }  \bigg)  
- {\kappa^2 \hp^2\over d(p^2) [d(p^2)^2  - \kappa^2 \hp^2]} \,
  \bigg(\hg_{\nu\lambda} - {\hp_\nu \hp_\lambda\over \hp^2} \bigg) \crn
&&\hskip 3cm     + {i\kappa \ep_{\nu\lambda\rho} p^\rho
   \over d(p^2)^2 - \kappa^2 \hp^2} ~~~. 
\label{inverse3}
\eeqn
In three dimensions, the propagator in the
Landau gauge  reduces to 
\beeq K^{-1}_{\nu\lambda}\Big|_{\rm 3-dim} =
- {1 \over d(p^2)^2  - \kappa^2 p^2} \, \bigg\{ d(p^2) 
  \bigg(g_{\nu\lambda} - {p_\nu p_\lambda\over p^2} \bigg) 
 - i\kappa \ep_{\nu\lambda\rho} p^\rho \bigg\}  
        \hskip .5cm {\rm for} ~ \alpha=0  
\label{inverse4}
\eneq

The propagator (\ref{inverse2}) can be decomposed into several pieces;
\beqn 
&&K^{-1}_{\mu\nu}\Big|_{\alpha=0} = \crn
&& - {1\over a} \bigg\{ {1\over {m_+ + m_-}} \bigg( 
      {1\over m_+}{1\over{p^2 - m_+^2}} + {1\over m_-}{1\over{p^2 - m_-^2}}
      \bigg) - {1\over m_3^2}{1\over{p^2}} \bigg\}
      (\hg_{\mu\nu}\hp^2 - \hp_\mu \hp_\nu) \cr
\myskip
&&  + {1\over a}{1\over m_+ + m_-} \bigg( {1\over p^2 - m_+^2}
    - {1\over p^2 - m_-^2} \bigg)\,  i \, {\kappa \over |\kappa|} \,
     \ep_{\mu\nu\rho} p^\rho \cr
\myskip
&&  - {1\over a}{1\over m_3^2}  \bigg( {1\over p^2 - m_3^2}
    - {1\over p^2} \bigg) \bigg( (g_{\mu\nu}p^2 - p_\mu p_\nu) -
    (\hg_{\mu\nu}\hp^2 - \hp_\mu \hp_\nu) \bigg) \cr
\myskip
&&  + {\kappa^2 (p^2 - \hp^2)\over (d^2-\kappa^2p^2)(d^2-\kappa^2 \hp^2)} 
    \bigg\{  {\kappa^2\over d} (\hg_{\mu\nu}\hp^2 - \hp_\mu \hp_\nu)
    - i\kappa \ep_{\mu\nu\rho} p^\rho \bigg\}
\label{inverse5}
\eeqn
Here
\beqn
m_\pm & =& m_\pm(v) = {1\over 2} \, \left\{ \sqrt{ {\kappa^2\over a^2} 
+ {4(ev)^2\over a}} \pm {|\kappa|\over a} \right\}  \crn
m_3^2 &=& m_+ m_- = {e^2 v^2\over a}  ~~~. 
\label{masses1}
\eeqn
There are several poles.  $m_\pm$ are the masses of physical gauge
bosons in three dimensional spacetime. $m_3$ is the mass of photons in
the extra-dimensional space.  The massless pole corresponds to the gauge
degree of freedom.   The last term in (\ref{inverse5}) behaves as
$1/p^5$ for large $p$.  It gives finite contributions which vanish
in the $n\go 3$ limit. 
It is instructive to write $m_\pm$ in terms of Higgs  
mass $m_\Hi$ and Chern-Simons mass $m_\CS$:
\beqn
&&m_\Hi = {ev\over \sqrt{a}} \next m_\CS = {|\kappa|\over a} \cr
\noalign{\kern 12pt}
&&m_\pm = {1\over 2} \bigg\{ \sqrt{m_\CS^2 + 4 m_\Hi^2} \pm m_\CS \bigg\} 
\label{masses2}
\eeqn

\sxn{One loop corrections in gauge theory}

The one-loop effective potential  can be evaluated easily.  
For $K^{\mu\nu}$ given in (\ref{inverse1}),
\beeq
\det K = (-1)^{n-3} A^{n-3} (A+ p^2 B) (A^2 - \kappa^2 \hp^2) 
\label{determinant}
\eneq
Hence $V_\eff^{(1)}(v)$ is, including the ghost contribution,
\beqn
&&V_\eff(v)^{\rm 1-loop} =
{\hbar\over 2} \int {d^np\over i(2\pi)^n} \,
\bigg\{ ~ \ln\,  \big[p^2 - m_1(v)^2\big] + \ln\,  \big[p^2 - m_2(v)^2\big] \cr
\noalign{\kern 12pt}
&& \hskip 3.cm
+ \ln\,  \big[\{ ap^2 - (ev)^2 \}^2 - \kappa^2 \hp^2\big]
+ (n-3) \ln\,  \big[ap^2 - (ev)^2\big] \cr
\noalign{\kern 14pt}
&& \hskip 3.cm 
+ \ln\,  {1\over \alpha} \, \big[p^2 - \alpha(ev)^2\big]  
- 2 \ln\,  \big[p^2 - \alpha (ev)^2\big]   ~ \bigg\}
\label{1loopV2}
\eeqn

Except for the third term, the integrals can be evaluated by the
standard formula (\ref{formula1}) in Appendix A.  The third term 
contains both $n$-dimensional $p^2$ and 3-dimensional $\hp^2$, and
needs extra care.  To evaluate it we consider
\beqn
F(x) = F(x;c,n) &=& \int {d^np\over i(2\pi)^n} \, 
\ln \, \big[ (p^2 - c^2)^2 - x \hp^2 \big] \cr
&=& \int {d^np\over (2\pi)^n} \, 
\ln \, \big[ (p^2 + c^2)^2 + x \hp^2 \big] ~~~.
\label{Ffunction1}
\eeqn
We write
\beqn
F(x) &=& F(0) + x \, F'(0) + \int_0^x dx_1 \int_0^{x_1} dx_2 \, F''(x_2) \cr
\noalign{\kern 10pt}
F(0) &=& \int {d^np\over (2\pi)^n} \,  \ln \,  (p^2 + c^2)^2 
 = -2 \, {\Gamma(-\onehalf n)\over (4\pi)^{n/2}} \, c^n\cr
\noalign{\kern 10pt}
F'(0) &=& \int {d^np\over (2\pi)^n} \, {\hp^2\over (p^2+c^2)^2} 
=  {3\over n} \int {d^np\over (2\pi)^n} \, {p^2\over (p^2+c^2)^2} 
= {3\over 2} {\Gamma(1-\onehalf n)\over (4\pi)^{n/2}} \, c^{n-2} \cr
\noalign{\kern 10pt}
F''(x) &=& - \int {d^np\over (2\pi)^n} \,
    {(\hp^2)^2 \over [(p^2+c^2)^2 + x \hp^2 ]^2} 
\label{Ffunction2}
\eeqn
The integral for $F''(x)$ is finite at $n=3$ so that we may set $n=3$:
\beqn
F''(x)_{n=3} &=& - \int {d^3 p\over (2\pi)^3} \,
    {(p^2)^2 \over [(p^2+c^2)^2 + x p^2 ]^2}  \cr
\noalign{\kern 10pt}
&=& - {1\over 2\pi^2} \int_0^\infty dp \,
    {p^6\over  [(p^2+c^2)^2 + x p^2 ]^2} \crn
&=& - {1\over 8\pi} {x+ 5 c^2\over (x+ 4c^2)^{3/2}}~~~.
\label{Ffunction3}
\eeqn
Here we have made use of  (\ref{3dIntegral2}).
Hence
\beqn
F(x)_{n=3} &=&
-2 \, {\Gamma(-\threehalves)\over (4\pi)^{3/2}} \, c^3
+ {3\over 2} {\Gamma(-\onehalf)\over (4\pi)^{3/2}} \, c \, x
- {1\over 8\pi} \int_0^x dx_1 \int_0^{x_1} dx_2 \,
 {x+ 5 c^2\over (x+ 4c^2)^{3/2}} \cr
\noalign{\kern 10pt}
&=& - {1\over 6\pi} (x+4c^2)^{1/2} (x+c^2) ~~.
\label{Ffunction4}
\eeqn
In other words, the integral $F(x)$ at $n=3$ is the same as the 
integral where $\hp^2$ is replaced by $p^2$ in (\ref{Ffunction1}).

Returning to (\ref{1loopV2}), we find
\beeq
V_\eff(v)^{\rm 1-loop}_{n=3} =
-{\hbar\over 12\pi} \bigg\{ m_1(v)^3 + 
m_+(v)^3 +  m_-(v)^3 + m_2(v)^3 - [\alpha (ev)^2]^{3/2} \bigg\}
\label{1loopV3}
\eneq
where
\beeq
m_+^3+m_-^3 = \sqrt{m_\CS^2 + 4 m_\Hi^2} \, \Big( m_\CS^2 + 
m_\Hi^2\Big) 
=\sqrt{{\kappa^2 \over a^2}+{4 e^2 v^2 \over a}}
                \Big( {\kappa^2 \over a^2}+{e^2 v^2 \over a}\Big).
\label{masses3}
\eneq
Imposing the renormalization conditions (\ref{rencond}), one finds
that
the effective potential at one loop is
\beqn
&&V_\eff(v)^{\rm 1~loop}
= {\nu\over 6!} \, v^6 + {\hbar\over 12\pi} {\kappa^3\over a^3} \, G(z) \cr
\noalign{\kern 10pt}
&&G(z) = 3 z - (1+4z)^{1/2} (1+z)
+ {2  (1 - 62 {\tilde M} +  240 {\tilde M}^2 )  \over
(1 + 4{\tilde M})^{11/2} } \, z^3 + 1\cr
&&z =  {a e^2 v^2 \over \kappa^2} ~~,~~ \tM = {a e^2 M\over \kappa^2} ~~.
\label{1loopVeff}
\eeqn
It was pointed out in  the previous paper that  one loop calculations do not
produce definitive results about symmetry breaking;  the minimum occurs  
at $v=0$ or $v\not= 0$,  depending on the choice of $M$. We need to go to 
two loop.
 
\bigskip

\sxn{Two loop corrections in gauge theory}

Relevant vertices for evaluating the  two loop effective potential are
\beqn
\L_{\rm cubic} &=& e A^\mu(\phi_1\dd_\mu\phi_2 - \phi_2\dd_\mu\phi_1)
                   + e^2 A_\mu^2\phi_1 v \cr
\noalign{\kern 5pt}
&& - {\lambda\over 3!} v\phi_1 (\phi_1^2 +\phi_2^2 )
   - {\nu\over 6!} (12v^3\phi_1\phi_2^2 + 20v^3\phi_1^3) \cr
\noalign{\kern 5pt}
&& -\alpha e^2v\phi_1 c^\dagger c ~~
    \label {cubic_Lagrangian}
\eeqn
and
\beqn
\L_{\rm quartic} &=& {1\over 2} e^2 A_\mu^2 (\phi_1^2 + \phi_2^2)
   -{\lambda\over 4!} (\phi_1^4 + \phi_2^4 + 2\phi_1^2\phi_2^2) \cr
&& - {\nu\over 6!} (15v^2\phi_1^4 + 18v^2\phi_1^2\phi_2^2 + 3v^2\phi_2^4)~~
   \label {quartic_Lagrangian}
\eeqn
The two loop effective potential is found by inserting
({\ref {cubic_Lagrangian}}) and ({\ref {quartic_Lagrangian}}) into 
({\ref {effectivePotential}}).  In the Landau gauge 
there are five types of diagrams to be evaluated.

\vskip 0.5 cm
\noindent (1) Two scalar loops 
%\hskip 3cm 
\begin{center}
\bpic(70,12)
\put(20,3){\circle{36}}
\put(56,3){\circle{36}}
\epic
\end{center}
\vskip 0.5 cm

The part of the interaction Lagrangian that produces this diagram is 
\beeq
\tilde\L_{q1} = -\alpha_1\phi_1^4 - \alpha_2\phi_2^4 - \alpha_3\phi_1^2
                  \phi_2^2
      \label {quartic_1}
\eneq 
where
\beqn
\alpha_1 &=& {\lambda\over 4!} + {\nu\over 4!} {v^2\over 2} \cr
\myskip
\alpha_2 &=& {\lambda\over 4!} + {\nu\over 5!} {v^2\over 2} \cr
\myskip
\alpha_3 &=& {{2\lambda}\over 4!} + {\nu\over 5!} 3v^2 ~~
     \label {constants}
\eeqn
The effective potential due to this is 
\beqn
V_{\eff(q1)} &=& 
     - i^6 \hbar^2\int {{d^np d^nq}\over {(2\pi)^{2n}}} \bigg\{
      {{3 \alpha_1}\over{(p^2 + m_1^2)(q^2 + m_1^2)}} +
      {{3 \alpha_2}\over{(p^2 + m_2^2)(q^2 + m_2^2)}} \cr
\myskip
&& \hskip 3.5 cm
    + {\alpha_3 \over {(p^2 + m_1^2)(q^2 + m_2^2)}} \bigg\} \cr
\noalign{\kern 15 pt}
&=& {\hbar^2 \mu^{2(n-3)} \over {(4\pi)^2}} 
  \Bigg\{ 3\left({\lambda\over 4!} + {{15\nu v^2} \over {6!}} \right)m_1^2 
 + 3\left( {\lambda\over {4!}} + {{3\nu v^2}\over {6!}} \right) m_2^2 \cr
\myskip
&& \hskip 3.5 cm
 + 2\left({\lambda\over{4!}} + {{9\nu v^2}\over {6!}} \right)m_1 m_2 \Bigg\}~.
\label{2loopVQ1}
\eeqn

\vskip 0.5 cm
\noindent (2) One scalar and one gauge loop
%\vskip 0. cm
%\hskip 2.5cm 

\begin{center}
\bpic(90,16)(0,13)  %(86, 16)(-10,13)
\PhotonArc(30,16)(18,0,360){0.8}{20}
\put(66,16){\circle{36}}
\epic 
\end{center}
\vspace{0.5cm}

For this diagram we have
\beeq
\tilde\L_{q2} = {1 \over 2} e^2 A_\mu^2 (\phi_1^2 + \phi_2^2)
      \label {quartic_2}
\eneq 
The effective potential due to this is 
\beqn
V_{\eff(q2)} &=& i^2 \hbar^2 {{e^2} \over 2} \int {{d^np d^nq}\over 
         {(2\pi)^{2n}}} i {K_\mu}^\mu(p)^{-1} \left[ {i\over {q^2 - m_1^2}} +  
	 {i\over {q^2 - m_1^2}} \right ] \cr
\myskip
&& - {e^2 \hbar^2 \over 2a} \int {{d^np d^nq}\over 
         {(2\pi)^{2n}}} \bigg[ {1\over q^2 + m_1^2} +
         {1\over q^2 + m_2^2} \bigg] \cr
\myskip
&& \times \bigg\{ \bigg[ {1\over {m_+ + m_-}} \bigg( 
      {1\over m_+}{1\over{p^2 + m_+^2}} + {1\over m_-}{1\over{p^2 + m_-^2}}
      \bigg) - {1\over m_3^2}{1\over p^2} \bigg] (2 \hp^2) \cr
\myskip
&& \hskip 2cm + {1\over m_3^2} \bigg[ {1\over{p^2 + m_3^2}} -
   {1\over p^2} \bigg] \big[(n-1)p^2 - 2\hp^2 \big] \bigg\} ~.
\eeqn
After the integration 
\beeq
V_{\eff}^{(q2)} = {e^2 \hbar^2  \mu^{2(n-3)} \over{16\pi^2}  a} 
{(m_1 + m_2) (m_+^2+m_-^2) \over m_+ + m_-} ~. 
\label{2loopVQ2}
\eneq

\vskip 0.5 cm
\noindent (3) $\theta$-shape diagram with pure scalar fields 
\begin{center}
\bpic(25, 0)
\put(18,3){\circle{36}}
\put(0,3){\line(1,0){36}}
\epic
\end{center}
\vskip 0.2 cm
This diagram is due to the following interaction Lagrangian :
\beqn
\tilde\L_{\rm c1} &=& - {\lambda\over 3!} v\phi_1 (\phi_1^2 + \phi_2^2 )
- {\nu\over 6!}  (12v^3\phi_1\phi_2^2 + 20v^3\phi_1^3) \cr
\myskip
&=& - \beta_1\phi_1^3 - \beta_2\phi_1\phi_2^2 
\eeqn
where 
\beqn
&& \beta_1 = {\lambda\over 3!} v +{\nu \over 36}v^3 \cr
\myskip
&& \beta_2 = {\lambda\over 3!} v + {\nu \over 60}v^3. 
\eeqn

The effective potential due to this is the same as the result obtained 
for the pure scalar theory given in Section 2.
\beqn
V_{\eff(c1)} &=& -{{i^8\hbar^2}\over 2} \int {{d^np d^nq}\over {(2\pi)^{2n}}}
           {1\over {(p+q)^2 + m_1^2}} 
	   \bigg\{ {{6\beta_1^2} \over {(p^2 + m_1^2)(q^2 + m_1^2)}} 
           +  {{2\beta_2^2} \over {(p^2 + m_2^2)(q^2 + m_2^2)}} \bigg\} \cr
\myskip
&=&  - \hbar^2    \Bigg[3\bigg({\lambda\over 3!} v + {\nu \over 36}v^3\bigg)^2 
 + \bigg({\lambda\over 3!}v + {\nu\over 60}v^3\bigg)^2 \Bigg] I^{\rm div} \cr
\myskip
&&      +{\hbar^2  \mu^{2(n-3)}\over {16\pi^2}}
         \bigg\{ 3 \bigg({\lambda\over 3!} v + {\nu \over 36}v^3\bigg)^2 
                      \ln {3m_1 \over \mu}
             + \bigg({\lambda\over 3!} v + {\nu \over 60}v^3\bigg)^2
 	              \ln {m_1 + 2m_2\over \mu} \bigg\} ~.
\label{2loopVC1}
\eeqn

\vskip 0.5 cm
\noindent (4) $\theta$-shape diagram with two scalar and one gauge propagators 
%\vskip 0. cm
\begin{center}
\bpic(70,16)(0,13)
\put(56,16){\circle{36}}
\Photon(38,16)(74,16){1}{10}
\epic
\end{center}
\vskip 0.0 cm 

For this diagram : 
\beeq
\tilde\L_{\rm c2} = e A^{\mu} (\phi_1 \d_\mu \phi_2 
			      -\phi_2 \d_\mu \phi_1)
\eneq  
The effective potential due to this is :
\beeq
V_{\eff(c2)} = {{i^6 e^2 \hbar^2}\over 2} \int {{d^np d^nq}
	    \over {(2\pi)^{2n}}} (p + 2q)^{\mu}
	    (p + 2q)^{\nu} K^{-1}_{\mu\nu}(p) \Delta_1(q)
	    \Delta_2(-p-q)
\label{Vc2a}
\eneq
where $K^{-1}_{\mu\nu}$, $\Delta_1$, and $\Delta_2$ denote $A_\mu$,
$\phi_1$  and
$\phi_2$ propagators, respectively.

Since
\beqn
(p + 2q)^{\mu} (p + 2q)^{\nu} (\hp^2\hg_{\mu\nu} - \hp_{\mu}\hp_{\nu})
     &=& 4 [\hp^2 \hq^2 - (\hp \cdot \hq)^2] \cr
\myskip
(p + 2q)^{\mu} (p + 2q)^{\nu} (p^2 g_{\mu\nu} - p_{\mu}p_{\nu})
     &=& 4 [p^2 q^2 - (p\cdot q)^2]
\eeqn
\ (\ref{Vc2a}) is reduced to three and $n$-dimensional integrals.
The effective 
potential becomes
\beqn
&&V_{\eff(c2)} = - {4e^2 \hbar^2\over 2} \int {{d^np d^nq}
	    \over {(2\pi)^{2n}}} {1\over{q^2 - m_1^2}}
	    {1\over{(p+q)^2 - m_2^2}} \cr
\myskip
&&\hskip .5cm \times \bigg\{ - {1\over a} \bigg[ {1\over {m_+ + m_-}} \bigg( 
   {1\over m_+}{1\over{p^2 - m_+^2}} + {1\over m_-}{1\over{p^2 - m_-^2}}
   \bigg) - {1\over m_3^2}{1\over{p^2}} \bigg]
   (\hp^2 \hq^2 - (\hp\cdot \hq)^2) \cr 
\myskip
&&\hskip 1cm  - {1\over a}{1\over m_3^2}  \bigg[ {1\over p^2 - m_3^2}
    - {1\over p^2} \bigg] \bigg[ (p^2 q^2 - (p\cdot q)^2) - 
        (\hp^2 \hq^2 - (\hp\cdot \hq)^2) \bigg] \cr
\myskip
&&\hskip 1cm 
  + {\kappa^2 (p^2 - \hp^2)\over (d^2-\kappa^2p^2)(d^2-\kappa^2 \hp^2)}
      {\kappa^2\over d} (\hp^2 \hq^2 - (\hp\cdot \hq)^2) \bigg\}  ~~~.
\eeqn
Employing (\ref{Kdef2}) and  (\ref{3dintegrals1}), one finds
\beqn
&&V_{\eff}^{(c2)}= {e^2 \hbar^2\over 2 a}  
\bigg[ 2(m_1^2 + m_2^2) - (m_+ + m_-)^2  + 3 m_3^2 \bigg] I^{\rm div} \cr
\myskip
&& \hskip 0.5 cm + {e^2 \hbar^2 \mu^{2(n-3)}\over 32 \pi^2 a}    \Bigg\{ 
\bigg [ m_1m_2 - {(m_1+m_2) \big\{ 2(m_1-m_2)^2 
   + m_+^2+m_-^2 \big\} \over m_+ + m_-}  \bigg ]  \crn
&& \hskip 0.5cm 
- {(m_1^2 - m_2^2)^2 \over m_3^2} \ln {m_1 + m_2\over \mu}   \cr
\myskip
&& \hskip 0.5cm 
 - \sum_{a = \pm}  {2m_a^2(m_1^2 + m_2^2)   - m_a^4  - (m_1^2 - m_2^2)^2
      \over m_a(m_+ + m_-)} \ln{m_a + m_1 + m_2 \over \mu} 
 - {5 \over 12} {\kappa^2 \over a^2}
     \Bigg\} ~. 
\label{2loopVC2}
\eeqn

\vskip 0.5 cm
\noindent (5) $\theta$-shape diagram with two gauge and one scalar propagators
%\vskip 0. cm
\begin{center}
\bpic(75,16)(0,13)
\PhotonArc(56,16)(18,0,360){0.8}{20}
\Line(38,16)(74,16)
\epic 
\end{center}
\vskip .2cm
The interaction Lagrangian is 
\beeq
\tilde\L_{c3} = e^2 A_\mu^2 \phi_1 v
\eneq 
and the corresponding  effective potential  is 
\beeq
V_{\eff(c3)} = i^6 \hbar^2 e^4 v^2 \int {{d^np d^nq}\over {(2\pi)^{2n}}} 
    K^{-1}(p)_{\mu \nu}  {K^{-1}}(q)^{\mu \nu} {1\over { (p+q)^2 - m_1^2}} ~~.
\label{Vc3a}
\eneq
Upon contracting the tensor indices between the gauge propagators  we have
\beqn
&& (\hp^2 \hg^{\mu\nu} - \hp^{\mu}\hp^{\nu})
   (\hq^2 \hg_{\mu\nu} - \hq_{\mu}\hq_{\nu}) 
       = \hp^2 \hq^2 + (\hp\cdot \hq)^2 \cr
\myskip
&& (i\kappa \epsilon^{\mu\nu\rho} p_{\rho})
   (i\kappa \epsilon_{\mu\nu\sigma} q^{\sigma})
       = -2 \kappa^2 (\hp\cdot \hq) \cr
\myskip
&&  \bigg[ (p^2 g^{\mu\nu} - p^{\mu}p^{\nu}) -
           (\hp^2 \hg^{\mu\nu} - \hp^{\mu}\hp^{\nu}) \bigg]
    \bigg[ (q^2 g_{\mu\nu} - q_{\mu}q_{\nu}) -
           (\hq^2 \hg_{\mu\nu} - \hq_{\mu}\hq_{\nu}) \bigg] \cr
\myskip
&& \hskip 2 cm 
   = (n-2)p^2 q^2 + (p\cdot q)^2 - 2(p^2\hq^2 + \hp^2 q^2) + 3\hp^2\hq^2
       - (\hp\cdot \hq)^2 \cr
\myskip
&&  (\hp^2 \hg^{\mu\nu} - \hp^{\mu}\hp^{\nu})
     \bigg[ (q^2 g_{\mu\nu} - q_{\mu}q_{\nu}) -
           (\hq^2 \hg_{\mu\nu} - \hq_{\mu}\hq_{\nu}) \bigg] \cr
\myskip
&& \hskip 2 cm
   = 2 \hp^2 ( q^2 - \hq^2) \cr
\myskip
&& \bigg[ {\kappa^2 \over d(p^2)} (\hp^2 \hg^{\mu\nu} - \hp^{\mu}\hp^{\nu})
         - i\kappa \epsilon^{\mu\nu\rho} p_{\rho} \bigg] 
   \bigg[ {\kappa^2 \over d(q^2)} (\hq^2 \hg_{\mu\nu} - \hq_{\mu}\hq_{\nu})
         - i\kappa \epsilon_{\mu\nu\sigma} q^{\sigma} \bigg] \cr
\myskip
&& \hskip 2 cm
   = {\kappa^4 \over d(p^2)d(q^2)} \bigg[\hp^2 \hq^2 + (\hp\cdot\hq)^2\bigg]
     - 2\kappa^2 (\hp \cdot\hq)  ~~.
\label{relation1}
\eeqn
Note that integrals involving the last term in the propagator 
$K^{-1}_{\mu\nu}$, (\ref{inverse5}),  vanish in the limit 
$n = 3$.  With (\ref{relation1}) the  effective potential
(\ref{Vc3a})  reduces to 
\beeq
V_{\eff(c3)} = \frac {i^6 \hbar^2 e^4 v^2}{a^2} \bigg [
               \tilde V_{c3a} + \tilde V_{c3b} + \tilde V_{c3c} +
               \tilde V_{c3d} \bigg]
\eneq  
where
\beqn
\tilde V_{c3a} &=&  \int {{d^np d^nq}\over {(2\pi)^{2n}}} 
                 \frac {\hp^2\hq^2 + (\hp\cdot\hq)^2} {(p+q)^2 - m_1^2}
	\bigg[{1\over m_{+} + m_{-}} \bigg( {1\over m_{+}}
              {1\over p^2 - m_{+}^2} +  {1\over m_{-}}
              {1\over p^2 - m_{-}^2} \bigg) -  {1\over m_{3}^2}
              {1\over p^2} \bigg] \cr
\myskip
&& \hskip 3 cm \times
	\bigg[{1\over m_{+} + m_{-}} \bigg( {1\over m_{+}}
              {1\over q^2 - m_{+}^2} +  {1\over m_{-}}
              {1\over q^2 - m_{-}^2} \bigg) -  {1\over m_{3}^2}
              {1\over q^2} \bigg] \cr
\myskip
\tilde V_{c3b} &=& - {\kappa^2 \over a^2} {1 \over (m_{+}^2 -  m_{-}^2)^2}
    \int {{d^np d^nq}\over {(2\pi)^{2n}}} 
    \frac {2\hp\cdot\hq} {(p+q)^2 - m_1^2} 	
    \bigg [  {1\over p^2 - m_{+}^2} -  {1\over p^2 - m_{-}^2} \bigg] \cr
\myskip
&&  \hskip 3 cm \times
	\bigg [  {1\over q^2 - m_{+}^2} -  {1\over q^2 - m_{-}^2} \bigg] \cr
\myskip
\tilde V_{c3c} &=& {1\over m_{3}^4} \int {{d^np d^nq}\over {(2\pi)^{2n}}} 
    \frac {(n-2)p^2q^2 + (p\cdot q)^2 - 2(p^2\hq^2 + \hp^2 q^2) + 3\hp^2\hq^2
       - (\hp\cdot\hq)^2} {(p+q)^2 - m_1^2} \cr
\myskip
&&  \hskip 3 cm \times
	\bigg [ {1\over p^2 - m_{3}^2} -  {1\over p^2} \bigg]
   	\bigg [ {1\over q^2 - m_{3}^2} -  {1\over q^2} \bigg] \cr
\myskip
\tilde V_{c3d} &=& {4\over m_3^2}  \int {{d^np d^nq}\over {(2\pi)^{2n}}}
   \frac {\hp^2 (q^2 - \hq^2)} {(p+q)^2 - m_1^2} 
	\bigg [ {1\over q^2 - m_{3}^2} -  {1\over q^2} \bigg]   \crn
&&  \hskip 3 cm \times
   \bigg[ {1\over m_{+} + m_{-}}  \bigg( {1\over m_{+}}
          {1\over p^2 - m_{+}^2} +  {1\over m_{-}}
          {1\over p^2 - m_{-}^2} \bigg) -  {1\over m_{3}^2}
          {1\over p^2} \bigg] 
\eeqn
Once again the integrals above can be evaluated, with the aid of 
(\ref{Kdef2}), (\ref{3dintegrals1}) and  (\ref{3dintegrals2}), to be
\beqn
&&V_{\eff}^{(c3)} = - 
  {3\hbar^2 e^4 v^2 \over 2 a^2} I^{\rm div}
 - {\hbar^2 e^4 v^2 \mu^{2(n-3)}\over 32\pi^2a^2} 
    \bigg[ - {2m_1 \over m_+ + m_-} -  {2m_1^2 + 12 m_3^2\over (m_+ + m_-)^2}
              + 3 \bigg] \cr
\myskip
&&\hskip .3cm  + {\hbar^2 e^4 v^2 \mu^{2(n-3)} \over 64\pi^2a^2} 
\Bigg\{  
    {2\big[(m_+ - m_-)^2 - m_1^2 \big]^2
       \over m_3^2(m_+ + m_-)^2}
              \ln {m_+ + m_- + m_1 \over \mu} 
    + {m_1^4\over m_3^4} \ln {m_1 \over \mu} \cr
\myskip
&& \hskip .3 cm
+ \sum_{a = \pm} \bigg[ {(4m_a^2 - m_1^2)^2 \over m_a^2(m_+ + m_-)^2}
              \ln {2m_a + m_1 \over \mu}
    - {2(m_a^2 - m_1^2)^2 \over m_3^2 m_a (m_+ + m_-)}
              \ln {m_a + m_1\over \mu} \bigg] \Bigg\} ~. \cr
\myskip
&& \hskip 5 cm
\label{2loopVC3}
\eeqn 

We have obtained the effective potential up to two loop order.
Two loop corrections yield divergent contributions.  Collecting all
divergent terms in (\ref{2loopVC1}), (\ref{2loopVC2}),
and (\ref{2loopVC3}), we see
\beqn
V_\eff(v)^{\rm div} &=& \hbar^2 \, C \, I^{\rm div} \crn
C &=&  - 3\bigg( {\lambda\over 3!} v + {\nu \over 36}v^3 \bigg)^2 
   - \bigg({\lambda\over 3!}v + {\nu\over 60}v^3\bigg)^2 \crn
&&
+ {2e^2\over a} \Big( m^2 + {\lambda\over 3} v^2 + {\nu\over 120} v^4
  \Big)
-{2e^4 v^2\over a^2} - {e^2\kappa^2\over 2a^3}  ~.
\label{divergent1}
\eeqn
These divergent terms are absorbed by counter
terms.  The renormalizability guarantees that divergent terms
are proportional to $v^0$, $v^2$, $v^4$, or $v^6$.  It is important 
to recognize that these counter terms are singular in $a$.  We shall
come back to this point when we discuss the Coleman-Weinberg limit in the
next section.
 
The effective potential in the $\MSbar$ scheme is obtained by simply dropping
divergent terms.  In the rest of this section
we investigate the behavior of the effective potential at small
and large $v$ analytically.
We investigate the global behavior of the  potential numerically in
Section 9.

In the $\MSbar$ scheme
\beeq
V_\eff (v)^\MSbar = V_\eff^{\rm (tree)}  + V_\eff^{({\rm 1-loop})}
	     + V_\eff^{({\rm 2-loop \,finite})} ~~. 
\label{totaleffpot}
\eneq
We are interested in the behavior of the effective potential in the 
massless limit defined by $m^2 = \lambda = 0$. 

To find the behavior of $V_\eff (v)^\MSbar$ for small $v$, we note
that
the expansion parameter $z$ is
\beeq
z = {a e^2 v^2\over \kappa^2} ~~~.
\label{expansion-parameter}
\eneq
The masses of the gauge bosons are expanded as
\beeq
m_{\pm} = \cases{
{\mybig \kappa\over \mybig a} (1+z-z^2+2z^3 + \cdots)\cr
{\mybig \kappa\over \mybig a} (z-z^2+2z^3 + \cdots)~~~. \cr}
\label{gauge-mass3}
\eneq
Up to two loop the small $v$ expansion is
\beeq
V_{\rm {small}}^\MSbar =    \sum_{n=1}^\infty C_{2n}v^{2n}
 +  \sum_{n=3}^\infty D_{2n}v^{2n} \ln v  ~~.
\label{Vsmalllimit1}
\eneq
The crucially important coefficient is $D_6$, which is
produced by logarithmic terms originating from $V_{c1}$, 
$V_{c2}$ and $V_{c3}$.
\beqn
V_{c1} &=& {\hbar^2 \over 32\pi^2} {7\over 675} \nu^2 v^6 \ln v
            + \cdots \cr
\myskip
V_{c2} &=& - {\hbar^2 \over 32\pi^2} \bigg[ {\nu \over 5} 
	{e^4 \over \kappa^2} - 2 {e^8\over \kappa^4}\bigg] v^6 \ln v 
   + \cdots \cr
\myskip
V_{c3} &=&  {\hbar^2 \over 32\pi^2} \bigg [
	14 {e^8 \over \kappa^4} - {\nu \over 6} 
	{e^4 \over \kappa^2} \bigg]  v^6 \ln v + \cdots ~~.
\eeqn
Combining all of these, we obtain 
\beeq
D_6 = {\hbar^2 \over 32 \pi^2} \bigg( 16 \, {e^8 \over \kappa^4}
	- {11\over 30} \, {\nu \, e^4 \over \kappa^2}
        + {7 \over 675} \,  \nu^2 \bigg)
\label{D_6}
\eneq

Since there are no $\ln v$,  $v^2 \ln v$, or $v^4 \ln v$ terms in 
({\ref{Vsmalllimit1}}), we may impose the same renormalization conditions 
\ ({\ref{rencond}}) as in pure scalar case.  With these
 renormalization conditions, 
the dominant behavior of the potential at small $v$ is given by
\beeq
V_{\rm small}(v) \sim   D_6 v^6  \ln {v \over {\sqrt M}}  ~~.
\label{Vsmalllimit2}
\eneq 
Since $D_6$ is always positive, we  conclude that the tree level minimum 
at $v=0$ has turned into a maximum. 

Next we turn to the behavior at large  $v$. Upon using the
inverse of the previous expansion parameter, the gauge boson masses are 
given by
\beeq
m_{\pm} = {e\,v \over \sqrt{a}} \pm {\kappa \over 2a} 
          + {\rm O}\bigg({1\over v} \bigg)~~.
\label{gauge-mass2}
\eneq
The dominant term for all the gauge boson masses are the
same. The
potential is parametrized to two loop as
\beeq
V_{\rm large}^\MSbar =     
  \sum_{n=0}^\infty F_{6-n} v^{6-n}  +
	\sum_{n=0}^\infty G_{6-n} v^{6-n} \ln v  ~.
\label{Vlargelimit1}
\eneq
Again, terms contributing to $G_6$ arise from $V_{c1}$, $V_{c2}$ and 
$V_{c3}$.  Looking at the logarithmic part term by term, we have 
\beqn
V_{c1} &=&  {\hbar^2 \over 32\pi^2} {7\over 675} \nu^2 v^6 \ln v
            + \cdots \cr
\myskip
V_{c2} &=&   O(v^4\ln\,v) +\cdots  \cr
\myskip
V_{c3} &=&   O(v^4\ln\,v) +\cdots .
\eeqn
The $v^6 \ln v$ terms in 
$V_{c2}$ and $V_{c3}$ exactly cancel.  

Note that the coefficients of $v^6 \ln
v$ terms in the above are independent of gauge couplings $a$, $e$ or 
$\kappa$. 
 $G_6$ turns out to be independent of any gauge couplings.
 $G_6$ is determined solely by the $V_{c1}$ term.
\beeq
G_6 = {7\hbar^2 \over 30\pi^2} {\nu^2 \over 6!} ~~~.
\label{G_6} 
\eneq

Similarly, one can check that $F_6$ term comes entirely from 
$V_{q1}$ and $V_{c1}$.    The above limit also corresponds to
expansion in small $\kappa$ for non-vanishing $a$.

The potential is positive at large $v$, thus establishing the
stability of the theory. Combining the result at small $v$, we conclude
the symmetry is spontaneously broken in the massless scalar theory.

$D_6$ and $G_6$ are independent of $a$.  This is no coincidence.
In general, a Feynman diagram for the effective potential at arbitrary
order in the $\MSbar$ scheme is written as a sum of terms of the form 
\beeq
\nu^{n_1} \, \lambda^{n_2} \, \Big( {e^2\over a} \Big)^{n_3} \,
% \Big( {\kappa\over a} \Big)^{n_4} \,
v^{n_4} \, f[ m_1, m_2, m_3, m_+,  m_-] 
\label{general1}
\eneq
where $f$ is a finite, well-defined function of various
$m_k(v, m, \lambda, \nu, a, e, \kappa)$'s.  This
follows
from the form of various vertices and the gauge field propagator
(\ref{inverse5}).  As we show in Section 10, the superficial degree
of divergence for a diagram involving at least one gauge field
propagator with no external legs is at most 2.  The last term 
 in (\ref{inverse5}) lowers the divergence degree by 3, and therefore
its contributions to the effective potential are finite and vanish
in the $n=3$ limit.  
This establishes the form (\ref{general1}).

The powers $n_1 \sim n_4$ are zero or positive
integers.  The equivalence relation (\ref{equivalence1}) 
implies that in the Landau gauge $\alpha=0$  the effective potential 
$V_\eff(v)$ can depend on gauge couplings only through $\kappa/e^2$ and
$\kappa/a$.

When higher loop corrections are included, the dominant part of the
effective potential at large $v$ takes the form
\beeq
V_{\rm large}^\MSbar = \sum_{k=1}^\infty G^{(k)}_6 v^6 (\ln v)^k 
 + \cdots \hskip 1cm  \hbox{for large  $v$.}
\label{LargeAsymptotic}
\eneq
The coefficients  $G^{(p)}_6$'s are dimensionless.  
$(\ln v)^k$ terms arise from logarithmically divergent integrals.
Furthermore for large $v$,
$m_1^2, m_2^2 \sim \nu v^4$ and $m_\pm^2, m_3^2 \sim e^2 v^2/a$
so that $(\ln v)^k$ terms do not depend on $\kappa$ at all.

The $\lambda$ dependence of $G^{(p)}_6$'s can appear only from vertices,
with the power $n_2 \ge 0$.
But  available dimensionless combinations $\lambda/m$ and
$a\lambda/e^2$
are singular in  the $m\go 0$ or $e\go 0$ limit.  
Since the $m\go 0$ limit is well defined for 
$v\not= 0$, $G^{(p)}_6$'s cannot depend on $\lambda/m$.  Similarly
the  theory with a vanishing  gauge coupling ($e=0$) is well defined.
This excludes the dependence on $\lambda$.
To summarize,
$G^{(p)}_6$'s depend on only $\nu$.

\bigskip

\noindent 
{\tt [Theorem]} ~ {\sl In the  scalar electrodynamics the 
coefficients  $G^{(p)}_6$'s defined in
(\ref{LargeAsymptotic}) for the effective potential at large
$v$ are independent of gauge couplings $a$, $e$, and $\kappa$ and
of $m$ and $\lambda$ to all order in perturbation theory.}

\bigskip

For small $v$ we consider the special case $m=\lambda=0$.  
The dominant part of the
effective potential at small $v$ is written as
\beeq
V_{\rm small}^\MSbar = \sum_{k=1}^\infty D^{(k)}_6 v^6 (\ln v)^k 
 + \cdots \hskip 1cm  \hbox{for small  $v$.}
\label{SmallAsymptotic}
\eneq
The coefficients  $D^{(k)}_6$'s are dimensionless, and therefore
can depend on $\nu$ and $e^2/\kappa$ only.

\bigskip

\noindent
{\tt [Theorem]} ~ {\sl In the massless scalar electrodynamics 
with $m=\lambda=0$ the 
coefficients  $D^{(k)}_6$'s defined in
(\ref{SmallAsymptotic}) for the effective potential at small
$v$ are independent of $a$ to all order in perturbation theory.}

\bigskip

\sxn{Coleman-Weinberg limit}

As explained earlier,  the Maxwell term is necessary to define the
theory  in the dimensional regularization scheme.  Without it the theory
loses renormalizability, as the gauge field propagator in
extra-dimensional space behaves badly at high momenta. 
The Coleman-Weinberg limit is  defined as the limit where
there is no dimensional parameter to start with. In this case, it corresponds
to the $a , m, \lambda \go 0$ limit.

The subtlety lies in the $a\go 0$ limit.  Loop corrections give rise
to  terms singular in $a$.  We shall see below that all these singular
terms are absorbed by counter terms at least to two loop.  The $a\go 0$ 
limit, the Coleman-Weinberg limit, is well defined after renormalization.

The effective potential is expressed in terms of $m_1(v)$, $m_2(v)$, 
$m_\pm(v)$, and $m_3(v)$.  Only the gauge boson masses depend on $a$.
The expansion in $a$ is thus equivalent to the expansion in $z$
defined in (\ref{expansion-parameter}), provided that $\kappa \not= 0$.  

Expanding $V_\eff(v)$ in $a$, one finds that with $m = \lambda = 0$,
\beqn
V_\eff(v) &=& V^{\rm tree} +  V^{\rm 1-loop} +  V^{\rm 2-loop} 
   +  V^{\rm c.t.} \crn
V^{\rm 1-loop} &=& -{\hbar \mu^{n-3}\over 12\pi} \Bigg\{
{\kappa^3\over a^3} + {\kappa e^2\over a^2} v^2 + {2e^6\over \kappa^3} v^6
+ {\rm O} (a) \Bigg\} \crn
 V^{\rm 2-loop} &=& 
	\hbar^2 C |_{m=\lambda=0}\,  I^{\rm div} 
	+  \sum_{n=0}^\infty A_{2n} \, a^{n-3} \, v^{2n} \crn
&& +  \sum_{n=0}^\infty B_{2n} \,  a^{n-3} \,  v^{2n} 
	\ln \bigg( \frac{k}{a} \bigg ) \, \, 
   +  \sum_{n=3}^\infty D_{2n}v^{2n} \ln v. 
\label{CWexpansion}
\eeqn

Here $C$ is given in (6.23) and 
\beqn
A_0 &=& - {\hbar \over 12 \pi} \kappa^3 
	- {5\over 6} {\hbar^2\over 64\pi^2} e^2 \, \kappa^2 
	-  {\hbar^2\over 64\pi^2}  e^2 \, \kappa^2 \ln \mu^2 \cr
\myskip
A_2 &=& - {\hbar \over 4 \pi} e^2 \kappa 
	+ {\hbar^2\over 16\pi^2} \sqrt{\nu \over 24} e^2 \kappa
	  ( 1 + {1\over \sqrt{5}}) 
	- {\hbar^2\over 32\pi^2} e^4 
	+ {\hbar^2\over 4\pi^2} e^4 \ln 2
        - {\hbar^2\over 16\pi^2} e^4 \ln \mu^2 \cr
\myskip
A_4 &=& - {\hbar^2\over 1280\pi^2} \nu e^2 
	+ {\hbar^2\over 640\pi^2}  \nu e^2 \ln \mu^2
	+ {\hbar^2\over 4\pi^2} {e^6 \over \kappa^2}
	- {\hbar^2\over 2\pi^2} {e^6 \over \kappa^2} \ln 2 \cr
\myskip
A_6 &=&  - {\hbar \over 576 \pi} {\nu^{3/2} \over \sqrt{6}}
	   ( 1 + {1\over 5\sqrt{5}})  
	 - {\hbar \over 6 \pi} {e^6 \over \kappa^3}
	 + {\hbar^2\over 16\pi^2} {\nu^2 \over 720} \bigg [
	    {39 \over 20} + {3 \over 4 \sqrt{5}} 
	   + {5\over 6} \ln \bigg( {3\over 8} \bigg) \cr
\myskip
&& \hskip 1cm
 	   + {1 \over 5} \ln \bigg( {1\over \sqrt{24}} + 
	     {1\over \sqrt{30}} \bigg) 
	   - {2 \over 5} \ln \bigg( 1  + {1\over \sqrt{5}} \bigg)
	   - {7 \over 160} \ln \bigg( {1\over \sqrt{24}} \bigg)
	\bigg] \cr
\myskip
&&      +  {427 \hbar^2\over 7680\pi^2}  {\nu^2 \over 720} \ln \nu
	-  {\hbar^2\over 768\pi^2}  {\nu^{3/2} \over \sqrt{24}}
	   {e^2 \over \kappa} \bigg( {52 \over 15} - {4 \over 15 \sqrt{5}}
	   \bigg ) \cr
\myskip
&&    + {\hbar^2\over 1536\pi^2} {\nu e^4 \over \kappa^2} 
	 \bigg( {7 \over 5} + {2 \over \sqrt{5}} \bigg )
      + {\hbar^2\over 32\pi^2} \sqrt{\nu \over 24} {e^6 \over \kappa^3}
	 \bigg( 16 + {3 \over \sqrt{5}} \bigg )
      - {13 \hbar^2\over 24\pi^2} {e^8 \over \kappa^4} \cr
\myskip
&&    + \bigg [ {7\hbar^2\over 4\pi^2} {e^8\over \kappa^4}
	      - {\hbar^2\over 192\pi^2} {\nu e^4\over \kappa^2}	
	      \bigg] \ln 2 \cr
\myskip 
&&    + \bigg [   {\hbar^2\over 32\pi^2} {e^8 \over \kappa^4}
	        - {\hbar^2\over 320\pi^2} {\nu e^4 \over \kappa^2} 
                + {\hbar^2\over 28800\pi^2} \nu^2 \bigg ] 
	    \ln \bigg({e^2\over \kappa} + \sqrt{\nu\over 24}
	             (1 + {1\over \sqrt{5}}) \bigg) \cr
\myskip
&&    + \bigg [   {\hbar^2\over 4\pi^2} {e^8 \over \kappa^4}
	        - {\hbar^2\over 192\pi^2} {\nu e^4 \over \kappa^2} 
                + {\hbar^2\over 36864\pi^2} \nu^2 \bigg ] 
	    \ln \bigg({2 e^2\over \kappa} + \sqrt{\nu\over 24} \bigg) \cr
\myskip
&&    + \bigg [ - {\hbar^2\over 32\pi^2} {e^8 \over \kappa^4}
	        + {\hbar^2\over 384\pi^2} {\nu e^4 \over \kappa^2} 
                - {\hbar^2\over 18432\pi^2} \nu^2 \bigg ] 
	    \ln \bigg({e^2\over \kappa} + \sqrt{\nu\over 24} \bigg) \cr
\myskip
B_0 &=& {\hbar^2\over 32\pi^2} e^2 \kappa^2 \cr
\myskip
B_2 &=& {\hbar^2\over 8\pi^2} e^4 \cr
\myskip
B_4 &=& - {\hbar^2\over 320\pi^2} \nu e^2 \cr
\myskip
B_6 &=& - {\hbar^2\over 4\pi^2} {e^8 \over \kappa^4}
	+ {11\hbar^2\over 1920\pi^2} {\nu e^4 \over \kappa^2} 
\eeqn
$D_6$ is given in (\ref{D_6}).  
 Notice that those terms singular in $a$ are of the form $v^n$ where
$n=0, 2, 4, 6$.  They are cancelled by counter terms.  The
renormalized theory has a well-defined $a\go 0$ limit. The coefficients
of $A_{2n}$ and $B_{2n}$ are related to $C_{2n}$ given in 
(\ref{Vsmalllimit1}) by
\beeq
C_{2n} = A_{2n} a^{n-3} + B_{2n}  a^{n-3} \ln \bigg( \frac{k}{a} \bigg )
\eneq

Let us consider the Coleman-Weinberg limit.  We take the $a\go 0$ limit 
with a given $v$. Adopting the renormalization condition (\ref{rencond}), 
one has
\beeq
V_\CW(v) = {\nu(M)\over 6!} v^6  
  +  D_6 v^6 \Big( \ln {v\over \sqrt{M}} - {49\over 20} \Big)  ~~~.
\label{CW2}
\eneq
Let us choose the 
renormalization point to be the location of the minimum $\sqrt{M}
=v_{\rm min}$. 
The condition for minimum  value of the effective potential is
\beeq
V'_{\rm CW}(v_{\rm min}) = v_{\rm min}^5 \bigg[ {\nu \over 5!} -
                      {137 \over 10} D_6 \bigg] = 0 ~~,
\eneq
from which it follows, with the aid of  ({\ref{D_6}}),
\beeq
\nu_{\rm CW} = \nu(v_{\rm min}^2) 
= {1\over 2} \bigg[ b_1 - \sqrt{b_1^2 - 4 b_2} \bigg]
\label{nu_CW}
\eneq
where
\beqn
b_1 &=& {495\over14} {e^4\over \kappa^2} + {1800\pi^2 \over 959\hbar^2} \cr
\myskip
b_2 &=&  {10800\over 7} {e^8\over \kappa^4}  ~~~. 
\eeqn
The Coleman-Weinberg limit potential is written as
\beeq
V_{\rm CW} = {\nu_{\rm CW} \over 1644} v^6 \bigg(
             \ln {v\over v_{\rm min}} - {1\over 6} \bigg)
\eneq

In the case 
$\nu = {\rm O} (e^8 / \kappa^4  )$, equation
({\ref{nu_CW}}) becomes
\beeq
\nu_{\rm CW} = {822 \hbar^2 \over \pi^2}{e^8 \over \kappa^4} 
\eneq
so that  the Coleman-Weinberg potential is 
\beeq
V_{\rm CW} =  {\hbar^2 \over 2 \pi^2} {e^8 \over \kappa^4} v^6 
            \bigg ( \ln {v \over v_{\rm min}} - {1 \over 6} \bigg)~~~. 
\label{VeffCW2}
\eneq 
The symmetry is spontaneously broken.  Dimensional transmutation takes
place.   The perturbation theory is
reliable as far as $e^2/\kappa$ is small.

\bigskip

\sxn{Pure Maxwell theory ($\kappa=0$)}

Another interesting limit is when the kinetic term for
the gauge fields is given by the Maxwell term only. This corresponds to taking 
$\kappa \go 0$ in the previous expression for the effective potential, 
keeping $a$ 
non-vanishing. Without loss of generality one can set $a=1$. This theory 
is parity preserving as opposed to Chern-Simons theory. At the tree, 
level in the limit of vanishing $m^2$ and $\lambda$, there is one 
dimensional parameter $e$. 
In the Landau gauge, the  gauge field propagator reduces to 
\beeq
K^{-1}_{\nu\lambda}\Big|_{\alpha=0}
= - {1\over (p^2 - e^2 v^2)} \bigg( g_{\nu\lambda} 
 -  {p_\nu p_\lambda \over p^2 }  \bigg)  
\eneq

In the rest of this section we set $m^2 = \lambda = 0$. 
The one loop contributions are simplified to 
\beeq
V_\eff^{\rm 1-loop} = - {\hbar \over {12 \pi}} \bigg [ 
\bigg\{ \Big({1\over 24}\Big)^{3\over 2} 
    + \Big({1\over 120} \Big)^{3\over 2} \bigg\} 
{\nu}^{3\over2}v^6 + 2 e^3 |v|^3 \bigg ] ~~. 
\eneq
There appears a $|v|^3$ correction to the tree level effective potential. 
The loop corrections take the form
\beqn
V_\eff(v)^{\rm loop} &=& 
    {\hbar^2 \over {32\pi^2}} v^2 \bigg [ (-e^2+{\nu v^2\over 40})^2 
-{139\nu^2 v^4\over 6!\cdot 60} \bigg ] 
% \crn &&\hskip 2cm    \times 
\bigg [ - {1\over {n-3}} - \gamma_E + 1 + \ln {4\pi} \bigg ] \cr
\myskip
&& + \bigg [ C_2 v^2 + C_3 v^3 + C_4 v^4 + C_5 v^5 + C_6 v^6 \bigg ] \cr
\myskip
&& + {\hbar^2 \over {16\pi^2}} v^2 \bigg [ (e^2-{\nu v^2\over 80})^2
 -{149\nu^2 v^4\over 6!\cdot 240} \bigg ] \ln {v^2 \over \mu} \cr
\myskip
&& + {\hbar^2 \over {64\pi^2}} v^2 \bigg [ (e^2-{\nu v^2\over 20})^2
 -{\nu^2 v^4\over 6!} \bigg ] 
\ln {1\over \mu}  \Bigg( v + \myfrac{ e \sqrt{120/\nu}} { 1  +\sqrt{5}} 
    \Bigg)^2  \cr  
\myskip
&& + {\hbar^2 \over {16\pi^2}} v^2 \bigg [ (e^2-{\nu v^2\over 96})^2
 +{\nu^2 v^4\over 9216 } \bigg ] 
  \ln { \big(v + {2 e \sqrt{24/\nu}  } \big)^2 \over \mu} \cr
\myskip
&& - {\hbar^2 \over {64\pi^2}} v^2 \bigg [ (e^2-{\nu v^2\over 24})^2
 \bigg ] \ln { \big(v + { e \sqrt{24/\nu} } \big)^2 \over \mu} 
\label{zerokappa}
\eeqn
where
\beqn
C_2 &=& {\hbar^2 \over {16\pi^2}} e^4 \ln{\nu \over 120}        
+ {\hbar^2 \over {32\pi^2}} e^4 \ln (1+\sqrt 5) \cr 
\myskip
C_3 &=&  -  {\hbar \over {6\pi}} e^3
        + {\hbar^2 \over {32\pi^2}} e^3 (1 +2\sqrt 5)\sqrt{\nu \over 120}\cr
\myskip
C_4 &=& {\hbar^2 \over {768\pi^2}} e^2 \bigg [
      2 \sqrt 5 +5 - 24\ln{\sqrt{\nu\over 120} (1+\sqrt 5)}
      \bigg ] \cr
\myskip
C_5 &=& - {\hbar^2 \over {960\pi^2}}e\nu(-1+\sqrt 5)\sqrt{\nu\over 120} \cr
\myskip
C_6 &=&   - {\hbar \over {12\pi}}\bigg[ \Big({1\over 24}\Big)^{3\over 2} 
+\Big({1\over 120} \Big)^{3\over 2}  \bigg] {\nu}^{3\over2} 
	+ {\hbar^2 \over {16\pi^2}}{\nu^2 \over 720}
	  \bigg ( {39 \over 20} + {3 \over {4 \sqrt{5}}} \bigg ) \cr
\myskip
&& + {\hbar^2 \over {32\pi^2}} {\nu^2 \over 36}
        \bigg [  {1 \over 12} \ln {9\nu\over 24 } + {1 \over 100} 
	\ln{\nu\over 120} (9+2\sqrt 5) \bigg]
\eeqn

There are terms of the form $v^2\ln v$ and $ v^4 \ln v$ so that the 
renormalization
conditions (2.31) cannot be imposed.  Both the second and 
fourth derivative of the effective 
potential must be evaluated at a non-vanishing value of the field.
Both $v^2 \ln v$ and $v^4 \ln v$ terms arise from $V_{c2}$ and $V_{c3}$ 
in (6.16) and (6.22), respectively. Their origin is traced back to the 
logarithmic terms in 
$m_{\pm}, m_1$ and/or $m_2$ appearing in both expressions. The $\ln v^2$ 
terms arise as
\beeq
\ln \bigg(\chi_1 m_+ + \chi_2 m_- + \chi_3 m_1 + \chi_4 m_2 \bigg)^2
= \ln v^2 + \ln \bigg((\chi_1 + \chi_2) \tmc 
	+ (\chi_3 \tma + \chi_4 \tmb) v \bigg)^2 .
\eneq
In this equation $\tmc$,$\tma$ and $\tmb$ are independent of $v$. In the 
limit of 
large $v$, the same expression as in (\ref{Vlargelimit1}) is obtained.

As the coefficient of the $v^2 \ln v$ term in (\ref{zerokappa}) is
positive, the $U(1)$ symmetry is spontaneously broken at two loop.

\sxn{Numerical analysis}

In Section 6, we have obtained the analytic expression for the effective
potential up to two loop order (see equation ({\ref{totaleffpot}})). 
To implement the renormalization conditions ({\ref{rencond}}), we need to 
calculate the sixth derivative of the effective potential, which is  
highly non-trivial.  Although we have $V_\eff(v)^\MSbar$ in the closed
form, each term in $V_\eff(v)^\MSbar$ leads to an extremely lengthy
expression when differentiated six times.  We have found that standard
symbolic manipulation aided by Mathematica or Maple is not of much help.

We adopt numerical evaluation to find the sixth derivative of 
$V_\eff(v)^\MSbar$ at finite $v$.  We have found that it is best to make 
use of the Cauchy integral formula.  First the  
effective potential is analytically continued to the complex $v$
plane.  We measure all dimensionful quantities in the unit of $e$ to
define   dimensionless quantities:
\beeq
\tilde V_{\rm eff} =  { V_{\rm eff} \over e^6} ,
\hskip 1cm
x = {v \over e} , 
\hskip 1 cm 
k = {\kappa \over e^2} ,
\hskip 1 cm
h^2 = {M \over e^2}
\label{dimensionless1}
\eneq
where M is the renormalization point.   Note
\beeq
\tilde V_{\rm eff} = \tilde V_{\rm eff}(x; \nu, k, a, h) ~~.
\label{dimensionlessV}
\eneq

The numerical analysis is further 
simplified by removing the pole and other terms proportional to $x^2$, 
$x^4$ and $x^6$ in $V^{\rm 2-loop}$  as those terms are completely
absorbed in the definition of counter terms.  After this procedure 
the effective potential takes 
the following form:
\beeq
\tilde V_{\rm eff} = {\nu \over 6!} x^6 + \tilde V_{\rm loop}
                     + \tilde V^{\rm counter-terms}
\label{Vnum}
\eneq
where $\tilde V^{\rm counter-terms} = \alpha_0 + {1\over2} \alpha_2 x^2
             + {1\over 4!} \alpha_4 x^4 + {1\over 6!} \alpha_6 x^6$.

The $n$-th $x$-derivative of the potential at $h$ is 
\beeq
\tilde V^{(n)}(h) = {n! \over 2\pi i} \int_C dz {\tilde V(z) \over
	 		(z - h)^{n+1} }
\label{Cauchy1}
\eneq
where the contour $C$ should not encircle any singularities of $\tilde V$.
The imaginary part of the above integral is zero within the numerical
precision. 

The counter terms are fixed by the renormalization
conditions. ({\ref{Vnum}}) can be rewritten as
\beqn
\tilde V_{\rm eff}(x; \nu, k, a, h) 
&=& {\nu \over 6!} x^6 + \tilde V_{\rm loop}(x)
		     - \tilde V_{\rm loop}(0)
		     - {1\over2} \tilde V_{\rm loop}^{(2)}(0) x^2 \cr
\myskip
&&  - {1\over 4!} \tilde V_{\rm loop}^{(4)}(0) x^4
    - {1\over 6!} \tilde V_{\rm loop}^{(6)}(h) x^6  ~~.
\label{numericequation}
\eeqn
With this definition $\nu \equiv \nu(h) = \tilde V_{\rm eff}^{(6)}(h)$. 
$\tilde V_{\rm loop}(0)$,   $\tilde V_{\rm loop}^{(2)}(0)$,
and $\tilde V_{\rm loop}^{(4)}(0)$ are evaluated analytically from
the small $v$ expansion in Section 6.  $\tilde V_{\rm loop}^{(6)}(h)$
is evaluated numerically by (\ref{Cauchy1}).

In fig.\ 2 the tree, 1-loop, and 2-loop effective potentials are 
 plotted for  typical values of  parameters. The importance
 of two loop corrections is recognized.
\begin{figure}[htb]
\epsfxsize=10cm  
\centerline{
\epsffile[47 192 474 501]{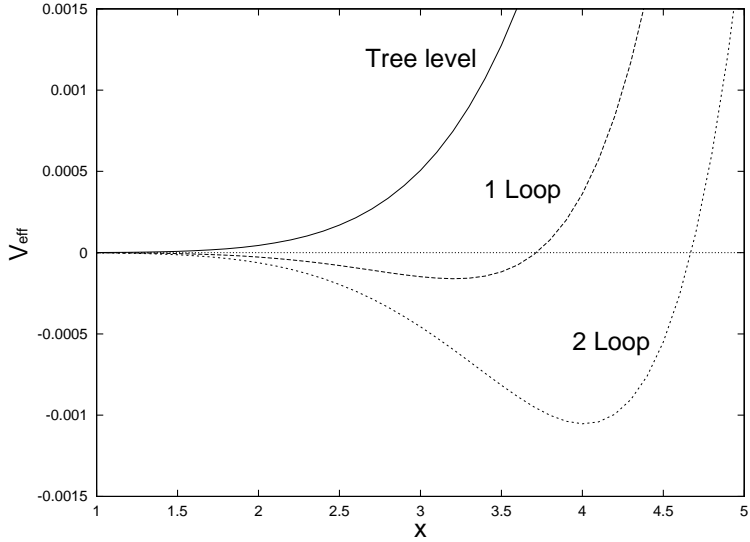}
}
%\vskip 1cm
\caption{Tree, 1-loop, and 2-loop effective potentials. 
 Plots for $a = 1$, $k = 20$, $\nu = 0.0005$, $h = 1$}
\label{fig:2}
%\vskip 0.2cm
\end{figure}
We also have depicted the effective potential for different values of 
parameters in fig.\ 3.

\begin{figure}[htb]
\epsfxsize=11cm  
\centerline{
\epsffile[47 192 474 501]{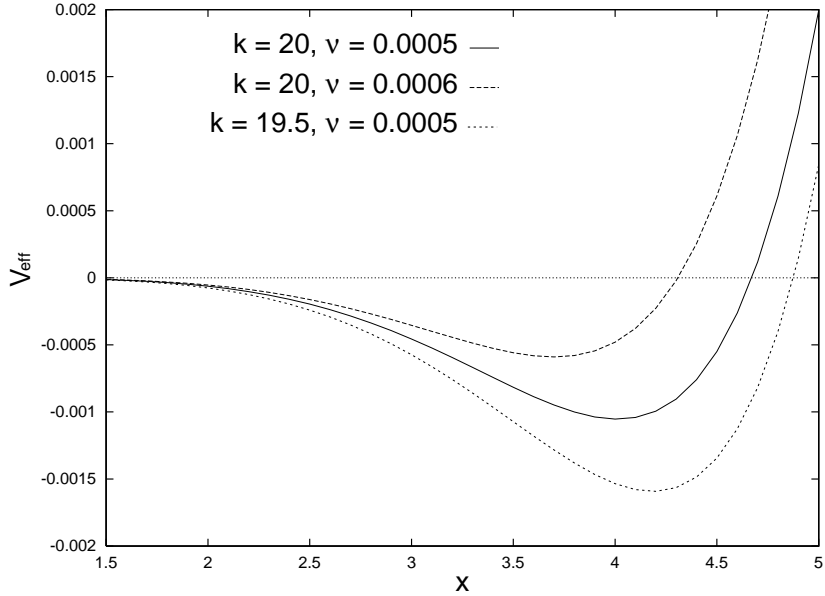}
}
%\vskip 1cm
\caption{The two loop effective potential plot for $a = 1$ and $h = 1$ using 
different values of $k$ and $\nu$}
\label{fig:3}
%\vskip 0.2cm
\end{figure}

Given $\nu$, $k$, $a$, and $h$, the potential is fixed.  It reaches a
minimum at $x=x_\min$.   $x_\min$ differs in general from $h$.
$\nu$ at the scale $x_\min$ is
\beeq
\nu(x_\min) = \tilde V_{\rm eff}^{(6)}(x_\min) 
\eneq
which differs from the initial $\nu=\nu(h)$.
Hence, the effective potential can be written as 
$\bar V_{\rm eff}(a, k, \nu(x_\min), x_\min; x)=
\tilde V_{\rm eff}(a, k, \nu, h; x)$. 

A detailed investigation of this yields some interesting properties. 
For example, a typical plot of $h \equiv h_{\rm in}$ vs
$x_\min \equiv h_{\rm out}$ is shown in fig.4. 
For particular values of parameters $\nu$ and $k$
the curve unexpectedly blows up in the region between $h_{\rm in} =$ 10 
and 30.   

\begin{figure}[htb]
\epsfxsize=10cm  
\centerline{
\epsffile[48 191 385 501]{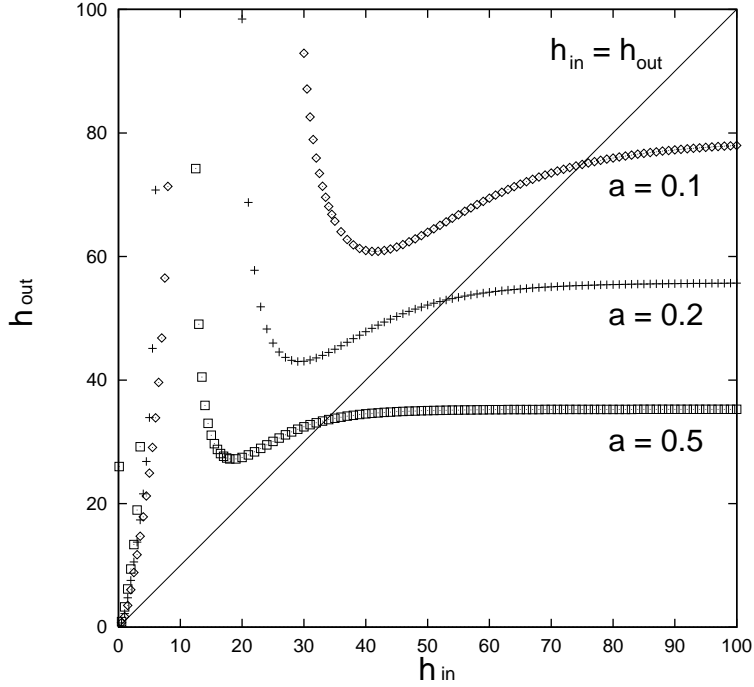}
}
%\vskip 1cm
\caption{$h_{\rm in}$ vs $h_{\rm out}$ plot for $k = 20$, $\nu$ = 0.0005
using various values of $a$}
\label{fig:4}
%\vskip 0.2cm
\end{figure}

\begin{figure}[htb]
\epsfxsize=10cm  
\centerline{
\epsffile[47 192 474 501]{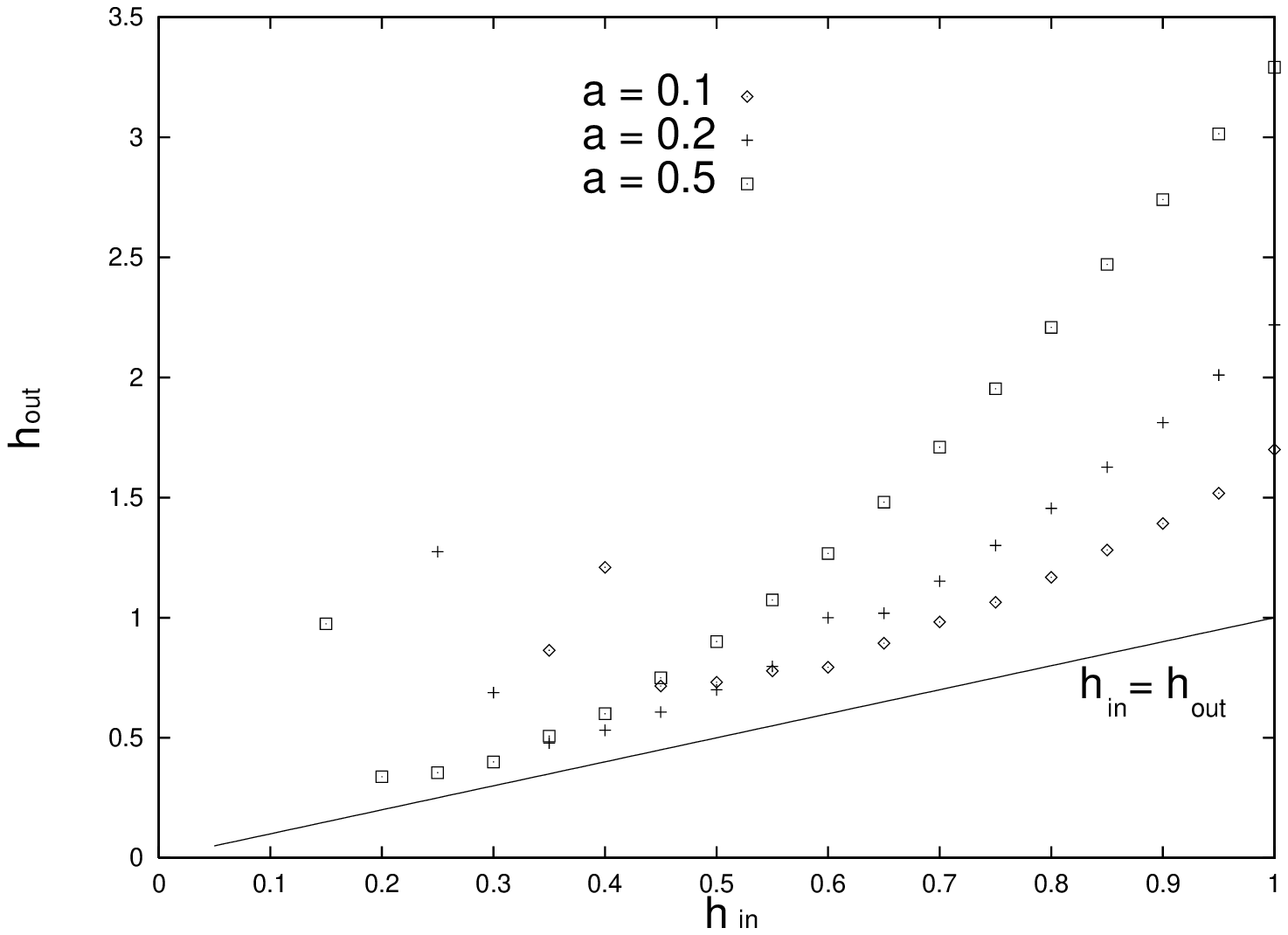}
}
%\vskip 1cm
\caption{Blown-up region of $h_{\rm in}$ vs $h_{\rm out}$ plot for small
$h_{\rm in}$ with $k = 20$ and $\nu = 0.0005$}
\label{fig:5}
%\vskip 0.2cm
\end{figure}

The region of small $h_{in}$ also shows some peculiar behavior 
which we have not been able to explain. We suspect that it could be due to 
the limitation of numerical evaluations or some unexplained phenomenon.
(See fig.\ \ref{fig:5}.)

We are also interested in  the Coleman-Weinberg limit of the 
potential. From the results of the previous
section, the effective potential in the Coleman-Weinberg limit is given by 
\beeq
\tilde V_{\rm CW} = {\hat \nu_{\rm CW} \over 1644} x^6 \bigg(
             \ln {x\over x_{\min}} - {1\over 6} \bigg)
\label{numeric_CW}
\eneq
where $\hat \nu_\CW= V_{\rm CW}^{(6)}(x_\min)$.  The potential is 
parametrized by two quantities $\hat \nu_\CW$ and $x_\min$.  The
location of the minimum is not determined by other parameters.  Instead it 
becomes an input parameter.  The limit $a\go 0$ must be taken with due
caution. 

As explained above, the input $h_{\rm in}=h$ and output $h_{\rm out}=x_\min$
are different in general.  As displayed in fig.\ 4, there is a fixed
point
value $h_{\rm out} = h_{\rm in}$ for given $\nu$, $k$, and $a$.  Take
this
value for $h$.  Then $x_\min=h$ and $\nu=\nu(x_\min)$.  In this
particular case $x_\min = x_\min(\nu, k, a)$.

Now we examine the $a$ dependence of $x_\min$.  The equivalence
relation (\ref{equivalence2}) implies that
\beeq
(a, k , e, \nu) \sim (a', k, e'=\sqrt{{a'\over a}} \, e, \nu) ~~.
\label{equivalence3}
\eneq
The two theories are the same so that the effective potential reach
the  minimum at the same $v$: $v_\min' = v_\min$.    In terms of the 
$x$ variable
\beeq
x_\min' = {v_\min' \over e'} = \sqrt{{a\over a'}} \, x_\min ~~.
\label{minimum1}
\eneq
In other words, if the $a\go 0$ limit is taken with  given $\nu$ and
$k$, then $x_\min \go \infty$, i.e.\ the Coleman-Weinberg limit is not
obtained.  This explains why the fixed point in fig.\ 4 moves to the
right as $a$ gets smaller.  The Coleman-Weinberg limit is not attained
because the expansion parameter $z$ in (\ref{expansion-parameter}) 
\beeq
z' = {a' {x_\min'}^2 \over k'^2} = {a x_\min^2\over k^2} = z
\label{expansion-parameter2}
\eneq
remains unchanged.

To get the Coleman-Weinberg limit one should not pick the fixed point
value for $h$.  One should choose $\nu$, $k$, and $h$ such that the 
expansion parameter at the minimum $z=ax_\min^2/k^2$  becomes small
when $a$ becomes small.  

Further in the Coleman-Weinberg limit $\nu(x_\min)$ and $k$ are
related  by (\ref{nu_CW}). This guides to the following procedure. 
Pick a value for $k$ and fix $\nu$ to be $\hat \nu_\CW(k)$.  Next
pick values for $h$ and $a$ such that $z <1$ and $h$ and $x_\min$ are
not terribly far apart.  With these $k$ and $h$, we make $a$ smaller 
to check if the potential approaches the Coleman-Weinberg limit.
At a given $a$ we compare the potential 
$\tilde V_\eff(x; \hat \nu_\CW, k, a, h)$ with the Coleman-Weinberg
potential (\ref{numeric_CW}) where $x_\min=x_\min(k,h,a)$  
is the location of the  minimum of $\tilde V_\eff(x)$.

In fig.\ 6 we displayed the result for $k=20$ and $h=1$.
For these values $\hat\nu_\CW=5.18\times 10^{-3}$.  One can see
that the two potentials get closer to each other as $a$ becomes smaller.

\begin{figure}[htb]
\epsfxsize=11cm  
\centerline{
\epsffile[47 192 474 501]{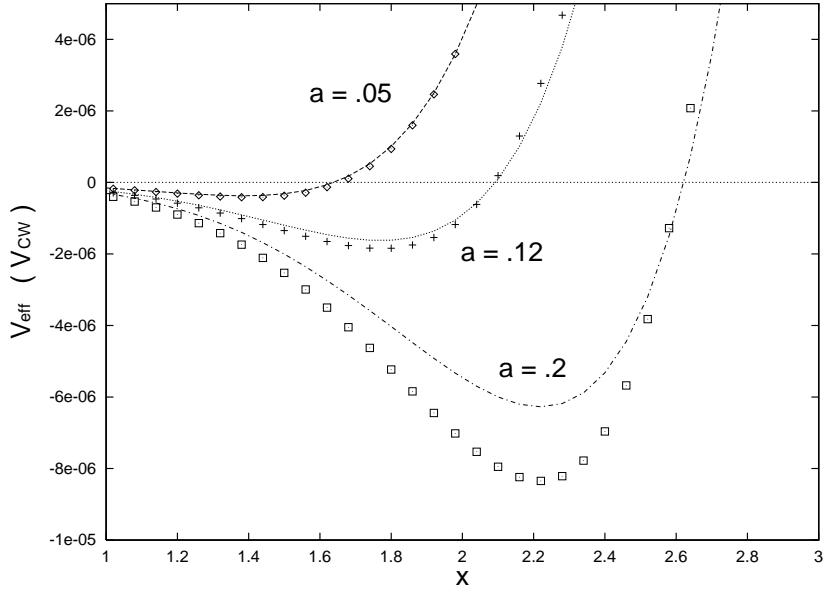}
}
%\vskip 1cm
\caption{Plots of $V_{\rm eff}$ and $V_{\rm CW}$ as a function of $x$ 
for different values of $a$. Points correspond to $V_{\rm eff}$. Lines 
correspond to $V_{\rm CW}$ determined by $x_{min}$ and $V_{\rm CW}(k)$ } 
\label{fig:6} %\vskip 0.2cm
\end{figure}

\sxn{Divergence structure }

It is helpful to understand the divergence structure of the theory
by examining the superficial degree of divergence in perturbation theory.
In doing so, one has to distinguish the $a= 0$  and $a\not=0$ case.
As we have observed in the preceding sections, the theory becomes
pathological if the perturbation theory is 
based on a free gauge field propagator with $a=0$. The 
Coleman-Weinberg theory has been defined by the limit $a\go 0$.

\leftline{(i) $\underline{\hbox{The case $a\not= 0$}}$}

The gauge field propagator is given by (\ref{inverse2}) - (\ref{inverse4}).
Notice that the propagator behaves as $1/p^2$ for large $p^2$:
\beeq
{\rm for} ~ a\not= 0 ,\quad K^{-1}_{\mu\nu} \sim {1\over p^2}
\quad  {\rm as} ~ p^2 \go \infty ~.
\label{asymptotic1}
\eneq
It is important that (\ref{asymptotic1}) is true in arbitrary dimensions $n$
and irrespective of whether $v=0$ or $v\not= 0$.
Hence it is sufficient to examine the superficial degree of divergence
in the unbroken theory $v=0$.  The ultra-violet behavior does not depend
on whether $v=0$ or $v\not= 0$.      

The Lagrangian (\ref{Lagrangian2}) yields
various vertices.    Let $V_4$, $V_6$, $V_{3A}$, $V_{4A}$, and $V_{3c}$ be 
the numbers of vertices $\phi^4$, $\phi^6$, $A\phi \dd \phi$, $A^2 \phi^2$, and
$\phi c^\dagger c$ in a given Feynman diagram $F$, respectively. 
We denote by $E$ and $I$ the number of
external and internal lines contained in $F$, respectively.  Then the number
of loop momenta $L$ is 
\beeq
L=I - V+1
\label{divergence1}
\eneq
where $V= V_4+V_6+V_{3A}+V_{4A}+V_{3c}$.

Since all propagators behave as $1/p^2$ and the vertex  $A\phi \dd \phi$
carries a derivative, the superficial degree of divergence in $n$ 
dimensions is
\beeq
\omega = n \, L - 2 I + V_{3A} ~~.
\label{divergence2}
\eneq
The topological identity $3(V_{3c}+V_{3A}) + 4(V_{4A}+V_4) + 6V_6 +E=2(E+I)$
gives
\beeq
I={1\over 2} \big\{ 3V_{3c} + 3V_{3A}  + 4V_{4A} +4V_4 + 6V_6 -E \big\} ~.
\label{divergence3}
\eneq
Combining (\ref{divergence1}) - (\ref{divergence3}), one finds
\beeq
\omega = 2(n-3) V_6 + (n-4) (V_4+V_{4A})  + {1\over 2} (n-4) V_{3A}
+ {1\over 2} (n-6) V_{3c}- {1\over 2} (n-2) E + n ~.
\label{DivergenceDegree1}
\eneq
In three dimensions $n=3$
\beqn
\omega &=& - V_4 - V_{4A} - {1\over 2} V_{3A}
- {3\over 2}  V_{3c}- {1\over 2}  E +3 \cr
L &=& 2 V_6 + V_4 + V_{4A}  + {1\over 2}  V_{3A}
+ {1\over 2}V_{3c}- {1\over 2}  E + 1 ~.
\label{DivergenceDegree2}
\eeqn

For propagators ($E=2$), $\omega = 2 - V_4 - V_{4A} - {1\over 2} V_{3A}
- {3\over 2}  V_{3c}$.  Divergent contributions to the wave function
renormalization for the scalar field, $Z_\phi$, come from only
$\nu^n$ terms.  The anomalous dimension is
\beeq
\gamma_\phi = \gamma_\phi(\nu) \quad \hbox{to all orders}~,
\label{gamma1}
\eneq
i.e.\ it does not depend on gauge couplings.
Since a diagram of a single loop is finite in the dimensional
regularization scheme, $\delta Z_\phi = \O (\nu^2)$.  In other words,
the anomalous dimension vanishes, $\gamma_\phi=0$, to the two loop order.
 The mass counter 
term for scalar fields is $\O(\lambda, e^2, \lambda^2, \lambda e^2, e^4)
\times \O (\nu^n)$.  To two loop 
$\delta m^2 = \O(\lambda^2, \lambda e^2, e^4)$.

Contributions to the gauge field propagator must satisfy $V_{4A} \ge 1$ or
$V_{3A} \ge 2$.  There is no divergent contribution proportional to
$A_\mu^2$ from the gauge invariance.  This implies that the coefficient of
$F_{\mu\nu} F^{\mu\nu}$ remains finite so that  the wave function
renormalization  factor $Z_A = 1$.  Hence the anomalous dimension
vanishes.
\beeq
\gamma_A = 0 \qquad  \hbox{to all orders}.
\label{gamma2}
\eneq
There could appear divergent
contributions to the Chern-Simons coefficient $\eps A_\mu F_{\nu\rho}$.  The
superficial degree of divergence for the Chern-Simons coefficient is 
$1- V_4 - V_{4A} - {1\over 2} V_{3A} - {3\over 2}  V_{3c}$. And hence
$\delta \kappa =\O(e^2 \nu^n)$ ($n\ge 2$). Since $V_{4A}\ge 1$ or 
$V_{3A}\ge 2$ , it vanishes at two loop.

Contributions to the coefficient of the vertex $A \phi \dd \phi$ have
$\omega = {1\over 2}- V_4 - V_{4A} - {1\over 2} V_{3A} - {3\over 2}  V_{3c}$.
A diagram must have at least one $e$, $V_{3A} \ge 1$.
Hence $\delta e = \O (e \nu^n)$ ($n \ge 2$).  $\delta\kappa$ and $\delta e$
are not independent.  The Coleman-Hill theorem \cite{Hill} ensures that 
$\delta(\kappa/e^2) = 0$.

Contributions to the vertex $\lambda \phi^4$ have 
$\omega = 1- V_4 - V_{4A} - {1\over 2} V_{3A} - {3\over 2}  V_{3c}$.
Hence $\delta\lambda = \O(\nu^n)$ ($n\ge 2$) or $\O(\lambda \nu^n, e^2\nu^n)$
($n\ge 1$).  Contributions to the vertex $\nu \phi^6$ have
$\omega = - V_4 - V_{4A} - {1\over 2} V_{3A} - {3\over 2}  V_{3c}$.
Since $\gamma_\phi=\gamma_\phi(\nu)$,  the beta function depends on only $\nu$:
\beeq
\beta_\nu = \beta_\nu(\nu) \qquad \hbox{to all orders}.
\label{beta1}
\eneq

\leftline{(ii) $\underline{\hbox{The case $a= 0$}}$}

We have to stress that the perturbation theory based on $a=0$ is 
inconsistent in the dimensional regularization supplemented with 
$\epsilon^{\mu \nu \rho}$ in (4.7). This is due to the 
behavior of the gauge field propagator at large momenta. The propagator
(\ref{inverse2}) behaves at large $p^2$ and $a=0$
\beeq
K^{-1}_{\nu\lambda} \sim
-{1\over (ev)^2} \Bigg[ 
   \bigg( g_{\nu\lambda} -  {p_\nu p_\lambda \over p^2 }  \bigg)  
 - \bigg(\hg_{\nu\lambda} - {\hp_\nu \hp_\lambda\over \hp^2} \bigg)   \Bigg]
   - {i  \ep_{\nu\lambda\rho} p^\rho
   \over  \kappa \hp^2} ~.
\label{asymptotic2}
\eneq
The first term does not vanish.  In particular, extra-dimensional 
components of $K^{-1}_{\nu\lambda}$ behaves as O($p^0$).  In other words,
higher loop diagrams with  many gauge field propagators behave very badly.
The theory in the dimensional regularization scheme loses the renormalizability
if $a$ is set to be zero in defining the perturbation
theory.  One consistent way to define the 
$a=0$ theory (the Coleman-Weinberg theory) is to take the limit $a\go 0$
after renormalization, which we have adopted in this paper.

Yet this does not entirely exclude the  possibility of defining a theory with
$a=0$.  One possibility is to stay in three dimensions, adopting the 
Pauli-Villars regularization method.  We have not checked the feasibility
of the Pauli-Villars regularization method beyond one loop, particularly
when the symmetry breaking takes place.  There is ambiguity in defining
regulator fields.

Here we add an argument concerning the divergence structure, assuming that
there exists a  regularization method defined entirely in three
dimensions, consistent to all orders at $a=0$. Should such a regularization 
scheme
exist, the gauge field propagator in the Landau gauge would be, as inferred
from (\ref{inverse5}),
\beeq
K^{-1}_{\nu\lambda}\big|_{\rm 3-dim} =
 {-1 \over  \kappa^2 p^2 - (ev)^4 } \, 
\bigg\{ - (ev)^2 \bigg(g_{\nu\lambda} - {p_\nu p_\lambda\over p^2} \bigg) 
 - i\kappa \ep_{\nu\lambda\rho} p^\rho \bigg\} ~. 
\label{inverse6}
\eneq
The asymptotic behavior is
\beeq
K^{-1}_{\nu\lambda} \sim {- i \ep_{\nu\lambda\rho} p^\rho\over \kappa p^2} 
= \O  \Big( {1\over p} \Big) ~.
\label{asymptotic3}
\eneq
We suppose that regulator fields have the same behavior.

Accepting (\ref{asymptotic3}), we derive the formula for the superficial
degree of divergence.  To distinguish gauge field propagators we introduce
the following notation.  The number of external  gauge, Fadeev-Popov
ghost, or
scalar fields is denoted by $E_A$, $E_c$,  or $E_\phi$, respectively. 
Similarly the number of internal  gauge, Faddeev-Popov ghost, or scalar
fields is  denoted by $I_A$, $I_c$, or $I_\phi$.  We have
$E=  E_A + E_c +E_\phi$ and $I = I_A + I_c +I_\phi$.

The identities (\ref{divergence1}) and (\ref{divergence3}) are still valid.
Because of (\ref{asymptotic3}), (\ref{divergence2}) is modified to
\beeq
\omega = n L - 2(I_\phi + I_c) - I_A + V_{3A} ~~.
\label{divergence4}
\eneq
The topological identity associated with gauge couplings is
$2V_{4A} + V_{3A} + E_A = 2(E_A + I_A)$, from which it follows that
\beeq
I_A = {1\over 2} \big( V_{3A} + 2 V_{4A} - E_A \big) ~~. 
\label{divergence5}
\eneq
Combining these, we have
\beeq
\omega = 3 - V_4 - {3\over 2} V_{3c} - {1\over 2} E_\phi
-E_A - {1\over 2} E_c ~~.
\label{DivergenceDegree3}
\eneq
The formula for $L$ remains the same as in (\ref{DivergenceDegree2}).
Notice that gauge couplings become marginal; the superficial degree of
divergence does not depend on $\nu$ or $e$.  (Recall that in the $a=0$ theory
$e$ appears only in the combination $\kappa/e^2$ which is dimensionless.)

The divergence structure is quite different. This time one would conclude
that $\gamma_\phi$, $\beta_\nu$,  $\beta_e$ and $\beta_\kappa$ are all
functions of $\nu$ and $e^2/\kappa$.  $\gamma_A=0$ still holds.  
We stress that this
conclusion is drawn on the assumption of the existence of a consistent
regularization method to all orders, which needs to be established. 

\sxn{Renormalization Group Analysis }

The RG equation for the effective potential in the $\MSbar$ scheme is
\beqn
\bigg [ \mu {\dd \over \dd \mu} + \beta_{\nu} {\dd \over \dd \nu} + 
         \beta_{\kappa} {\dd \over \dd \kappa} 
       + {\beta_{e^2}\over 2} {\dd \over \dd {e^2}}
       + \beta_{a} {\dd \over \dd a} 
       + \beta_{m^2} {\dd \over \dd m^2} 
       + \beta_{\lambda} {\dd \over \dd \lambda} 
     + \beta_{\Lambda} {\dd \over \dd {\Lambda}} 
     - \gamma_{\phi} v  {\dd \over \dd v} \bigg ] &&\cr
\myskip
 \times
	V(v; \nu, m^2, \lambda, \kappa, e^2, a,\Lambda, \mu)^\MSbar = 0 &&
\label{RGeqn}
\eeqn
where various $\beta$ functions are given by (\ref{beta-fun1}) and
\beeq
\beta_\kappa = \mu{\dd\over \dd\mu} \, \kappa ~~,~~
\beta_{e^2} =  \mu{\dd\over \dd\mu} \, e^2 ~~,
\beta_a =  \mu{\dd\over \dd\mu} \, a ~~.
\label{beta-fun4} 
\eneq
The renormalization for $a$ is the same as the wave function
renormalization for
$A_\mu$.  Although the result (\ref{gamma2}) implies $\beta_a =0$,
we have kept the $\beta_a$ term in (\ref{RGeqn}) to show a useful relation
below. Note that up to O($\hbar^2$), $\gamma_{\phi} = 0$.   

In Section 2 we have shown that beta functions in pure scalar theory
can be determined from the renormalization group equation for the 
effective potential.  We employ the same technique to find
beta functions in the gauge theory.

At O($\hbar$), (\ref{RGeqn}) yields 
\beeq
\beta_{m^2}^{(1)} \, {v^2 \over 2} 
+ \beta_{\lambda}^{(1)} \, {v^4 \over 4!}
+ \beta_{\nu}^{(1)} \, {v^6 \over 6!} 
+ \beta_{\Lambda}^{(1)}  = 0
\label{RGeqn1}
\eneq
so that
\beeq
\beta_{m^2}^{(1)} = \beta_{\lambda}^{(1)} = \beta_{\nu}^{(1)} =
\beta_{\Lambda}^{(1)} = 0
\eneq

At O($\hbar^2$),  Eq.\ (\ref{RGeqn}) becomes :
\beqn
&& - {\hbar^2 \over {16 \pi^2}} \bigg [
      \bigg( {\lambda \over 6} v + {\nu \over 60} v^3 \bigg)^2 
  + 3 \bigg( {\lambda \over 6} v + {\nu \over 36} v^3 \bigg)^2 \bigg ] 	\cr
\myskip
&& + {e^2 \hbar^2 \over {32 \pi^2 \, a}} \bigg [
     4 \, m^2 + {4\over 3} \lambda \, v^2 + {\nu \over 10} v^4 
   - {\kappa^2 \over a^2} - {e^2 v^2 \over a} \bigg ] \cr
\myskip
&& - {3 e^4 \hbar^2 \over {32 \pi^2 \, a^2}} \, v^2  
+ \beta_{m^2}^{(2)} {v^2 \over 2} + \beta_{\lambda}^{(2)} {v^4 \over 4!}
+ \beta_{\nu}^{(2)} {v^6 \over 6!} + \beta_{\Lambda}^{(2)}  \cr
\myskip
&& -  {\hbar \over 12 \pi} 
 \bigg( \myfrac{\kappa^2}{a^2} + \myfrac{4 e^2 v^2}{a} \bigg)^{-1/2}
\Bigg[
 \myfrac{\beta_{\kappa}^{(1)}}{\kappa} 
    \myfrac{3\kappa^2}{a^2} 
    \bigg( \myfrac{\kappa^2}{a^2} + \myfrac{3e^2  v^2}{a} \bigg)
+\myfrac{\beta_{e^2}^{(1)}}{e^2} 
    \myfrac{3 e^2v^2}{a} 
    \bigg( \myfrac{\kappa^2}{a^2} + \myfrac{2e^2 v^2}{a}  \bigg) \cr
\myskip
&&\hskip 6cm 
-\myfrac{\beta_a^{(1)}}{a}
     \bigg( \myfrac{3\kappa^4}{a^4} + \myfrac{12 \kappa^2 e^2 v^2}{a^3}
           + \myfrac{6 e^4 v^4}{a^2} \bigg)
\Bigg] =0 ~.
\label{RGeqn2}
\eeqn
The above equation is quite complicated but must be satisfied for arbitrary
$v$.  Since the last term contains a square root, it must vanish
identically.  It follows immediately that
\beeq
{\beta_{\kappa}^{(1)} \over \kappa} 
   = {\beta_{e^2}^{(1)} \over e^2}
   = {\beta_{a}^{(1)} \over a} ~~.
\label{beta-fun5}
\eneq
Upon making use of $\beta_a=0$, one concludes that $\beta_{\kappa}^{(1)}
=\beta_{e^2}^{(1)} =0$.

Then the rest of the equation becomes
\beqn
{\rm O}(v^0) :&&  
{e^2 \hbar^2 \over {32 \pi^2 \, a}} \bigg [
     4 \, m^2 - {\kappa^2 \over a^2} \bigg ] 
+ \beta_{\Lambda}^{(2)}  = 0 \cr
\noalign{\kern 20 pt}
{\rm O}(v^2)  :&&  
- {\hbar^2 \over 144 \pi^2} \lambda^2 
+ {\hbar^2 \over 32 \pi^2} \bigg ({4\over 3} \lambda {e^2 \over a}
     - {e^4 \over a^2} \bigg)
- {3 \hbar^2 \over 32 \pi^2} {e^4 \over a^2} 
+ {1\over 2}\beta_{m^2}^{(2)}         = 0 \cr
\noalign{\kern 20 pt}
{\rm O}(v^4) :&&  
- {\hbar^2 \over 480 \pi^2} \lambda \, \nu 
+ {\hbar^2 \over 320 \pi^2} {e^2 \over a} \, \nu
+ {\beta_{\lambda}^{(2)} \over 24}  = 0 \cr
\noalign{\kern 20 pt}
{\rm O}(v^6) :&& 
   -  {\hbar^2 \over 16 \pi^2}{ 7 \over 2700} \nu^2 
   +  {\beta_{\nu}^{(2)} \over 720} 
   = 0 ~~. 
\label{beta-fun6}
\eeqn 
It follows that
\beqn
&& \beta_{\Lambda}^{(2)} = {e^2 \hbar^2 \over {32 \pi^2 \, a}} \bigg (
     4 \, m^2 - {\kappa^2 \over a^2} \bigg)  \cr
\myskip
&&  \beta_{m^2}^{(2)} = {\hbar^2 \over 72 \pi^2} \lambda^2 
-  {\hbar^2 \over 12 \pi^2} \lambda {e^2 \over a}
+  {\hbar^2 \over 4 \pi^2} {e^4 \over a^2} \cr
\myskip
&& \beta_{\lambda}^{(2)} = {\hbar^2 \over 20 \pi^2} \lambda \, \nu 
- {3 \hbar^2 \over 40 \pi^2} {e^2 \over a} \, \nu  \cr
\myskip
&&\beta_{\nu}^{(2)} = {7 \hbar^2 \over 60 \pi^2} \nu^2 ~~.
\label{beta-fun7}
\eeqn
The last relation for $\beta_\nu$ confirms the result (\ref{beta1}) at two
loop.

The beta functions in the $\MSbar$ scheme are singular the $a \go 0$ 
limit. The  the renormalization 
group equation for the Coleman-Weinberg potential (\ref{CW2}) is more
involved than naively expected.  Eq.\ (\ref{RGeqn}) is for the
effective potential in the $\MSbar$ scheme before renormalization.
These two are related by
\beqn
&&V_\CW(v; \nu, \kappa/e^2, M)  \crn
&&\hskip 1.cm = \lim_{a\go 0}
\Bigg\{  V_\MSbar(v) - V_\MSbar(0) - {v^2\over 2} V_\MSbar^{(2)}(0)
 - {v^4\over 4!} V_\MSbar^{(4)}(0) 
 - {v^6\over 6!} V_\MSbar^{(6)}(M^{1/2})        \Bigg\} \cr
\myskip
&&V_\MSbar(v) \equiv 
    V(v; \nu, m^2=0, \lambda=0, \kappa, e^2, a,\Lambda, \mu)^\MSbar~~~. 
\label{CW3}
\eeqn
The subtraction terms give additional contributions to
 the renormalization group equation.   

\sxn{Conclusion}

We have examined the Maxwell-Chern-Simons gauge theory with complex
scalar fields with the most general renormalizable interactions  at two loop.
The effective potential for the scalar fields was obtained in the 
closed form in dimensional regularization scheme.  In the massless
scalar theory the $\phi^6$ coupling constant $\nu$ cannot be
renormalized at $\phi=0$ as two loop corrections yield terms of the 
form $\phi^6 \ln \phi$.  Evaluation of the sixth derivative of the 
effective potential at finite $\phi$ is a formidable task, which we have
done by numerical method.  The renormalized effective potential
for general couplings was evaluated numerically.

We have found that two loop corrections are decisive to determine the
phase.  The $U(1)$ symmetry is spontaneously broken in the massless
theory ($m=\lambda=0$) by radiative corrections.  In particular, 
in the Coleman-Weinberg limit in which the Maxwell term is absent for
gauge fields, the dimensional transmutation takes place at two loop.

From the effective potential we have also determined beta functions
for various couplings. Two loop results confirm the general theorem
that the beta function $\beta_\nu$ is independent of gauge couplings
and a function of $\nu$ only.

Here we would like to stress again that the regularization of the theory
is a delicate matter. The Maxwell term (with the
coefficient $a$) must be introduced to have
improved ultraviolet behavior of the gauge field propagator in the dimensional
regularization.   We have demonstrated  that only after
renormalization  one can take the limit $a\go 0$.  Counter terms are
singular in $a$.  The perturbation theory defined with $a=0$ is 
inconsistent in the dimensional regularization scheme.  It is not 
renormalizable.  

Avdeev, Grigoryev and Kazakov have studied the  pure Chern-Simons
theory  coupled to scalar matter to find beta functions differing from
ours.\cite {Avdeev}  They evaluated diagrams in three dimensional
space to eliminate all $\eps$ tensors, and then extend and perform  momentum
integrals in $n$ dimensions to define ``dimensional regularization''.
This is incorrect.  Everything must be defined in $n$ dimensions
first.  This is the source of the discrepancy.

In the absence of  the Maxwell term one of the gauge
degrees of freedom becomes infinitely massive.  However, it cannot
be completely discarded.   It gives  nontrivial cancellations and the
beta function for the scalar field becomes independent of the gauge 
couplings. 

The $U(1)$ symmetry is spontaneously broken 
at the two loop
level.  Our results can be extended to supersymmetric 
self-dual Chern-Simons theory.  As it was pointed out by \cite{Lee},
in the $ N=2$ and  3 supersymmetric models 
the scaling symmetry broken at two loop may be restored  quantum mechanically.
 
\vskip 1cm 

{\small \baselineskip=12pt  
\leftline{\bf Acknowledgement}
The authors would like thank to A. Vainstein, G. Dunne, R. Pisarski for 
useful discussions. This work was supported in part  by 
the U.S.\ Department of Energy under contracts DE-AC02-83ER-40105.
}
\vskip 15pt

\axn{ Some Useful Formulas}

In this appendix we collect $n$ dimensional integrals which 
we have made use of in the paper. See also \cite{Collins}. 

In Minkowski space we have
\beqn
\int{d^nk\over i(2\pi)^n} \, \ln (m^2 - k^2) 
&=& - {\Gamma(-\onehalf n)\over (4\pi)^{n/2}} \, m^n\cr
\noalign{\kern 7pt}
\int{d^nk\over i(2\pi)^n} \, {1\over (m^2 - k^2)^\alpha} 
&=& + {\Gamma(\alpha -\onehalf n)\over (4\pi)^{n/2}\Gamma(\alpha)} ~ 
   (m^2)^{(n/2) - \alpha}\cr
\noalign{\kern 7pt}
\int{d^nk\over i(2\pi)^n} \, {k^\mu k^\nu\over (m^2 - k^2)^\alpha} 
&=& - {\Gamma(\alpha -1 -\onehalf n)\over (4\pi)^{n/2}\Gamma(\alpha)} ~ 
  {g^{\mu\nu}\over 2}  \, (m^2)^{(n/2)+1  - \alpha}
\label{formula1}
\eeqn
Also we have
\beqn
\int{d^nk\over i(2\pi)^n} \, {1\over (- k^2)^\alpha} 
&=& 0
\eeqn

Similarly in Euclidean space
\beqn
J_1[p,\alpha] &=&
\int{d^nk\over (2\pi)^n} \, {1\over [k^2+m^2 +x(2pk +p^2)]^\alpha} \cr
\noalign{\kern 8pt}
&=& + {\Gamma(\alpha -\onehalf n)\over (4\pi)^{n/2}\Gamma(\alpha)} ~ 
   [p^2x(1-x) + m^2]^{(n/2) - \alpha}\cr
\noalign{\kern 12pt}
J_2^\mu[p,\alpha] &=&
\int{d^nk\over (2\pi)^n} \, {k^\mu \over [k^2+m^2 +x(2pk +p^2)]^\alpha} 
= -x p^\mu \, J_1[p,\alpha] \cr
\noalign{\kern 12pt}
J_3^{\mu\nu}[p,\alpha] &=&
\int{d^nk\over (2\pi)^n} \, {k^\mu k^\nu\over [k^2+m^2 +x(2pk 
+p^2)]^\alpha}\cr 
\noalign{\kern 8pt}
&=& x^2 p^\mu p^\nu J_1[p,\alpha] 
 + {\delta^{\mu\nu}\over 2(\alpha-1)} \, J_1[p,\alpha-1] \cr
\noalign{\kern 8pt}
&=& x^2 p^\mu p^\nu {\Gamma(\alpha -\onehalf n)\over (4\pi)^{n/2}\Gamma(\alpha)}
~  [p^2x(1-x) + m^2]^{(n/2) - \alpha}\cr
&&\hskip .5cm + \onehalf \delta^{\mu\nu} \, {\Gamma(\alpha-1 -\onehalf n)\over
(4\pi)^{n/2}\Gamma(\alpha)} ~  [p^2x(1-x) + m^2]^{(n/2) - \alpha+1}.
\label{formula2}
\eeqn

\noindent
Below we give the integrals in the regularized form
\beqn
&&\int {d^3p\over (2\pi)^3} \, {1\over p^2+a^2}
 = -{a\over 4\pi} \cr
\noalign{\kern 14pt}
&&\int {d^3p\over (2\pi)^3} \, {1\over (p^2+a^2)(p^2+b^2)}
  = {1\over 4\pi} {1\over a+b}\cr
\noalign{\kern 12pt}
&&\int {d^3p\over (2\pi)^3} \, {1\over (p^2+a^2)(p^2+b^2)(p^2+c^2)}
= -{1\over 4\pi} \Big\{ af(a;b,c)+ bf(b;c,a) +cf(c;a,b) \Big\} \cr
\noalign{\kern 14pt}
&&\int {d^3p\over (2\pi)^3} \, {p^2\over (p^2+a^2)(p^2+b^2)(p^2+c^2)}
= {1\over 4\pi} \Big\{ a^3f(a;b,c)+ b^3f(b;c,a) +c^3f(c;a,b)  \Big\} \cr
\noalign{\kern 12pt}   
&& \hskip 4cm f(a;b,c) = {1\over (a^2-b^2)(a^2-c^2)}. 
\label{3dIntegral1}
\eeqn
As an application we have
\beeq
\int_{-\infty}^\infty dp \, {p^6\over (p^2+a^2)^2 (p^2 + b^2)^2}
= {\pi\over 2} ~{a^2 + 3ab+b^2\over (a+b)^3} ~~.
\label{3dIntegral2}
\eneq

\axn{Two loop integrals}

A basic two loop diagram yields
\beqn
I( m_1,m_2,m_3;n) &\equiv& \int \frac{d^nqd^nk}{(2\pi)^{2n}}
   \frac{1}{\left[ (q+k)^2+m_1^2\right] (q^2+m_2^2)(k^2+m_3^2)}\crn
&=& I(m_2, m_1,m_3;n) \quad {\rm etc.} \crn
&=&
{\mu^{2(n-3)}\over 32 \pi^2}  \bigg\{ - {1\over n-3} - \gamma_E + 1 
- \ln {(m_1+m_2+m_3)^2\over 4\pi \mu^2} \bigg\} ~.
\label{2loopIntegral1}
\eeqn
This result was first derived in \cite{Kajantie}.

To show this, we note
\beqn
I &=& \mu^{2(n-3)} {\Gamma(3-n)\over (4\pi)^3} \int_0^1 dxdy \,
{1\over \sqrt{x(1-x)y}} \bigg[ {y\over x(1-x)} \bigg]^{(3-n)/2}\cr
\noalign{\kern 10pt}
&&\hskip 2.cm \times
\Bigg[ {1\over 4\pi\mu^2} 
\bigg\{
 y \bigg( {\mybig m_1^2\over \mybig 1-x}
              +{\mybig m_2^2\over\mybig x} \bigg)   +(1-y) m_3^2
 \bigg\} \Bigg]^{n-3}\cr
\noalign{\kern 10pt}
&=&  {\mu^{2(n-3)}\over (4\pi)^3} \int_0^1 {dxdy \over \sqrt{x(1-x)y}}
\Bigg\{ -{1\over n-3} - \gamma_E + {1\over 2} \ln {y\over x(1-x)} \cr
\noalign{\kern 10pt}
&&\hskip 1cm   - \ln \Bigg[ {1\over 4\pi\mu^2}   \bigg\{
 y \bigg( {\mybig m_1^2\over \mybig 1-x}
 +{\mybig m_2^2\over\mybig x} \bigg)   +(1-y) m_3^2 \bigg\} \Bigg] \Bigg\}
\label{2loopIntegral2}
\eeqn
In the $y$ integral, we have
\beeq
\ln \Big\{ p(x) y + q \Big\} \next
p(x) = {m_1^2\over 1-x} + {m_2^2\over x} - m_3^2 \next q = m_3^2>0 .
\eneq
$p(x)$ ($0<x<1$) reaches a minimum at $x=m_2/(m_1+m_2)$.  Its value is
$p_{\rm min}= (m_1+m_2)^2 - m_3^2$.  Hence, making use of
\[
\int_0^1 dy \, {\ln (py+q)\over \sqrt{y} }
= 2 \ln (p+q) - 4 + 4 \sqrt{{q\over p}} \tan^{-1} \sqrt{{p\over q}} ~~,
\]
we find, for $m_1+m_2 \ge m_3$, 
\beqn
I&=&{\mu^{2(n-3)}\over 32\pi^2} 
\Big( -\frac{1}{n-3}-\gamma_E + 2\ln 2 + 1 \Big) \cr
\myskip
&&\hskip 0cm - {\mu^{2(n-3)}\over 32\pi^3} \int_0^1 {dx\over \sqrt{x(1-x)}}
\ln \Bigg[ {1\over 4\pi\mu^2}   
 \bigg( {\mybig m_1^2\over \mybig 1-x} + \myfrac{m_2^2}{x} \bigg)  \Bigg] \cr
\myskip
&&\hskip 0cm 
- {\mu^{2(n-3)}\over 16\pi^3} \int_0^1 {dx\over \sqrt{x(1-x)}}
{\tan^{-1} \bigg(
\myfrac{1}{1-x} \myfrac{m_1^2}{m_3^2} 
  + \myfrac{1}{x} \myfrac{m_2^2}{m_3^2}  -1\bigg)^{1/2}\over 
\bigg( \myfrac{1}{1-x} \myfrac{m_1^2}{m_3^2} 
  + \myfrac{1}{x} \myfrac{m_2^2}{m_3^2}  -1 \bigg)^{1/2}}
\eeqn
The second term is evaluated as
\beeq
\int_0^1 {dx\over \sqrt{x(1-x)}}
\ln \Bigg[ {1\over 4\pi\mu^2}   
 \bigg( {\mybig m_1^2\over \mybig 1-x} + \myfrac{m_2^2}{x} \bigg)  \Bigg]
= \pi\ln {(m_1 + m_2)^2\over \pi\mu^2} ~~.
\eneq
Hence
\beqn
I_{m_1+m_2\ge m_3} &=&{\mu^{2(n-3)}\over 32\pi^2} 
\Bigg\{ -\frac{1}{n-3} -  \ln {(m_1 + m_2)^2\over 4\pi\mu^2}
- \gamma_E + 1 
  + f \Big( {m_1^2\over m_3^2}, {m_2^2\over m_3^2} \Big) \Bigg\} \cr
\myskip
f(a,b) &=&  
- {2\over \pi} \int_0^1 {dx\over \sqrt{x(1-x)}}
{\tan^{-1} \bigg(
\myfrac{a}{1-x}  + \myfrac{b}{x} -1\bigg)^{1/2}\over 
   \bigg( \myfrac{a}{1-x}  + \myfrac{b}{x} -1 \bigg)^{1/2}} ~~.
\eeqn

As a special case we have
\beeq
I(m_1,m_2,0;n) = {\mu^{2(n-3)}\over 32\pi^2} 
\Bigg\{ -\frac{1}{n-3} -  \ln {(m_1 + m_2)^2\over 4\pi\mu^2}
- \gamma_E + 1 \Bigg\} ~~,
\label{2loopIntegral3}
\eneq
which can be obtained  directly from (\ref{2loopIntegral2}), too.  Since the 
divergent term in (\ref{2loopIntegral3}) is independent of $m_j$, one can write
\beeq
I(m_1,m_2,m_3;n) = I(m_1,m_2,0;n) + 
\int_0^{m_3} dm_3 \, {\dd\over \dd m_3} I(m_1,m_2,m_3;n)\Big|_{n=3} 
 + {\rm O}(n-3)~~.
\label{2loopIntegral4}
\eneq

Now we evaluate
\beqn
&&{\dd\over \dd m_3} I(m_1,m_2,m_3;n)\Big|_{n=3} \crn
&&=-2m_3 \int \frac{d^3qd^3k}{(2\pi)^{6}}
   \frac{1}{\left[ (q+k)^2+m_1^2\right]^2 (q^2+m_2^2)(k^2+m_3^2)}\crn
&&=-{m_3\over 4\pi^4} \int_0^\infty dqdk\int_{-1}^1 d(\cos\theta) \,
{q^2k^2\over [q^2+k^2+2qk\cos\theta +m_3^2]^2 (q^2+m_1^2) (k^2+m_2^2)} \crn
&&={m_3\over 8\pi^4} \int_0^\infty dqdk\,
{qk\over (q^2+m_1^2) (k^2+m_2^2)}  \Bigg\{
{1\over (q+k)^2+m_3} -{1\over (q-k)^2+m_3^2} \Bigg\} \crn
&&={m_3\over 32\pi^4} \int_{-\infty}^\infty dqdk\,
{qk\over (q^2+m_1^2) (k^2+m_2^2)}  \Bigg\{
{1\over (q+k)^2+m_3^2} -{1\over (q-k)^2+m_3^2} \Bigg\} ~~.
\eeqn 
Making use of the residue theorem, one finds
\beqn
&&{\dd\over \dd m_3} I(m_1,m_2,m_3;n)\Big|_{n=3} \crn
&&={m_3\over 16\pi^3} \int_{-\infty}^\infty dk\, {ik\over k^2+m_2^2}
\Bigg\{ {1\over 2} \Bigg({1\over (k+im_1)^2+m_3^2} 
 -{1\over (k-im_1)^2+m_3^2} \Bigg)  \crn
&&\hskip 3cm +{-k+im_3\over (k-im_3)^2+m_1^2}{1\over 2im_3}
-{k+im_3\over (k+im_3)^2+m_1^2}{1\over 2im_3} \Bigg\} \crn
&&={1\over 64\pi^3} \int dk \, {k\over k^2+m_2^2} \crn
&&\hskip .3cm \Bigg\{
{1\over k+im_1-im_3} - {1\over k+im_1+im_3}  
-{1\over k-im_1-im_3} + {1\over k-im_1+im_3} \crn
&&\hskip .5cm 
-{1\over k-im_3-im_1} - {1\over k-im_3+im_1} 
-{1\over k+im_3-im_1} - {1\over k+im_3+im_1} \Bigg\}\crn
&&={-1\over 16\pi^3} \int dk \, {k^2\over (k^2+m_2^2) [k^2+(m_1+m_3)^2]} \crn 
&&= -{1\over 16\pi^2} {1\over m_1+m_2+m_3}  ~~~.
\label{2loopIntegral5}
\eeqn
Hence eq. (\ref{2loopIntegral4}) becomes
\beeq
I(m_1,m_2,m_3;n)=  -{\mu^{2(n-3)}\over 16\pi^2} \ln{m_1+m_2+m_3\over m_1+m_2}
+ I(m_1,m_2,0;n) + {\rm O}(n-3) 
\eneq
which leads, with (\ref{2loopIntegral3}),  to (\ref{2loopIntegral1}).

\axn{More loop integrals}

In the course of evaluating loop corrections, we would typically encounter 
integrals involving both three and $n$-dimensional momenta. These 
integrals can be reduced into a basic Euclidean integral of the following 
form :
\beqn
J^{\mu\nu\rho\sigma} &=& \int \frac{d^npd^nq}{(2\pi)^{2n}}
      \frac{p^{\mu} p^{\nu} q^{\rho} q^{\sigma}}
           {(p^2 + m_1^2) (q^2 + m_2^2) [(p+q)^2 + m_3^2]} \cr
\myskip
&=& A g^{\mu\nu} g^{\rho\sigma} + B \bigg( g^{\mu\rho} g^{\nu\sigma}
     + g^{\mu\sigma} g^{\nu\rho} \bigg)
\label{Jintegral1}
\eeqn
Upon contracting (\ref{Jintegral1}) with $g_{\mu\nu} g_{\rho\sigma}$
and $ g_{\mu\rho} g_{\nu\sigma}$ independently, we obtain 
\beqn
&& n^2 \, A + 2 \, n \, B = K_1 \cr
\myskip
&& n \, A + (n^2 + n) \, B = K_2
\label{Jintegral2}
\eeqn
where
\beqn
K_1 &=& \int \frac{d^npd^nq}{(2\pi)^{2n}} \frac{p^2 \, q^2} 
        {(p^2 + m_1^2) (q^2 + m_2^2) [(p+q)^2 + m_3^2]} \cr
\myskip
K_2 &=& \int \frac{d^npd^nq}{(2\pi)^{2n}} \frac{(p\cdot q)^2} 
        {(p^2 + m_1^2) (q^2 + m_2^2) [(p+q)^2 + m_3^2]}
\label{Kdef1}
\eeqn
Both $K_1$ and $K_2$ are expressed in terms of the $I$ integral given in
the previous section.
\beqn
K_1 &=&  - {\mu^{2(n-3)} \over 16\pi^2} (m_1^3 + m_2^3) m_3
         + m_1^2 + m_2^2 I(m_1, m_2, m_3) \cr
\myskip
K_2 &=& {\mu^{2(n-3)} \over 64\pi^2}
        \bigg\{ m_3^2 \bigg[ m_1 m_3 + m_2 m_3 - m_1 m_2 \bigg] 
           - m_3 \bigg[ 3 m_1^3 + 3 m_2^3 + m_1^2 m_2 + m_2^2 m_1 \bigg] \cr
\myskip
&& \hskip 0.5 cm   + (m_1^2 + m_2^2) m_1 m_2 \bigg\} 
 + {1\over 4} \bigg[ 4 m_1^2 m_2^2 - (m_1^2 + m_2^2 - m_3^2)^2 \bigg]
     I(m_1, m_2, m_3)
\label{Kdef2}
\eeqn 
Solving (\ref{Jintegral2}), we may write (\ref{Jintegral1}) in terms
of $K_1$ and $K_2$.
\beeq
J^{\mu\nu\rho\sigma} 
= \frac {(n+1) K_1 - 2 K_2} {n(n - 1)(n+2)} g^{\mu\nu} g^{\rho\sigma} 
   +  \frac {n K_2 - K_1} {n(n - 1)(n+2)}
     \bigg( g^{\mu\rho} g^{\nu\sigma} + g^{\mu\sigma} g^{\nu\rho} \bigg)
\label{Jintegral3}
\eneq
Making use of (\ref{Jintegral3}), one finds 
\beqn
&&\int \frac{d^npd^nq}{(2\pi)^{2n}} \frac{\hp^2 \, \hq^2} 
        {(p^2 + m_1^2) (q^2 + m_2^2) [(p+q)^2 + m_3^2]} 
= \frac {(9n + 3) K_1 + (6n - 18) K_2} {n(n - 1)(n+2)} \cr
\myskip
&&\int \frac{d^npd^nq}{(2\pi)^{2n}} \frac{(\hp\cdot \hq)^2} 
        {(p^2 + m_1^2) (q^2 + m_2^2) [(p+q)^2 + m_3^2]}
= \frac {(3n -9) K_1 + (12n - 6) K_2} {n(n - 1)(n+2)} \cr
\myskip
&&\int \frac{d^npd^nq}{(2\pi)^{2n}} \frac{\hp^2 \, \hq^2 - (\hp\cdot\hq)^2} 
        {(p^2 + m_1^2) (q^2 + m_2^2) [(p+q)^2 + m_3^2]}
= \frac {6}{n(n-1)} (K_1 - K_2) \cr
\myskip
&&\int \frac{d^npd^nq}{(2\pi)^{2n}} \frac{\hp^2 \, \hq^2 + (\hp\cdot\hq)^2} 
        {(p^2 + m_1^2) (q^2 + m_2^2) [(p+q)^2 + m_3^2]}
= \frac {6(2n - 1) K_1 + 6(3n - 4) K_2} {n(n - 1)(n+2)} \cr
\myskip
&&\int \frac{d^npd^nq}{(2\pi)^{2n}} \frac{p^2 \, \hq^2} 
        {(p^2 + m_1^2) (q^2 + m_2^2) [(p+q)^2 + m_3^2]} 
= {3 \over n} K_1 
\label{3dintegrals1}
\eeqn
One must be careful when solving the above integrals for $n = 3$. The 
pole terms in $K_1$ and $K_2$ give finite contributions upon being 
 multiplied by $(n-3)$.  

Other useful 2-loop integrals are :
\beqn
&&K_3 = \int \frac{d^npd^nq}{(2\pi)^{2n}} \frac{2 \, p . q} 
        {(p^2 + m_1^2) (q^2 + m_2^2) [(p+q)^2 + m_3^2]} \cr
\myskip
&& \hskip 0.5 cm
= {\mu^{2(n-3)} \over 16\pi^2} (m_1 m_2 - m_2 m_3 - m_1 m_3)
	+ (m_1^2 + m_2^2 - m_3^2) I(m_1, m_2, m_3) \cr
\myskip
&&\int \frac{d^npd^nq}{(2\pi)^{2n}} \frac{2 \, \hp . \hq} 
        {(p^2 + m_1^2) (q^2 + m_2^2) [(p+q)^2 + m_3^2]}
= {3\over n} K_3 \cr
\myskip
&& \hskip 0.5cm
=  {\mu^{2(n-3)} \over 16\pi^2} (m_1 m_2 - m_2 m_3 - m_1 m_3)
        +  {\mu^{2(n-3)} \over 96\pi^2} (m_1^2 + m_2^2 - m_3^2) ~ \cr
\myskip
&& \hskip 2cm
	+ (m_1^2 + m_2^2 - m_3^2) I(m_1, m_2, m_3) + O(n-3).
\label{3dintegrals2}
\eeqn

\vskip 1cm
\newpage
\baselineskip=16pt  
\leftline{\bf References}

\renewenvironment{thebibliography}[1]
	{\begin{list}{[$\,$\arabic{enumi}$\,$]}  % {\arabic{enumi}.}
	{\usecounter{enumi}\setlength{\parsep}{0pt}
	 \setlength{\itemsep}{0pt}  \renewcommand{\baselinestretch}{1.2}
         \settowidth
	{\labelwidth}{#1 ~ ~}\sloppy}}{\end{list}}

\end{document}